\documentclass[10pt]{article}
\usepackage{amsmath,amssymb}
\usepackage{graphicx}
\usepackage{hyperref}
\DeclareGraphicsExtensions{.eps,.bmp,.wmf,.jpg,.pdf}
\numberwithin{equation}{section}
\def\be{\begin{equation}}
\def\ee{\end{equation}}

\def\bea{\begin{eqnarray}}
\def\eea{\end{eqnarray}}
\title{\textbf{Inflation in a scalar-vector-tensor theory}}
\author{A. Oliveros\thanks{alexanderoliveros@mail.uniatlantico.edu.co}\,\, and
Cristhian J. Rodr\'iguez\thanks{cjoserodriguez@mail.uniatlantico.edu.co}\\
Programa de F\'isica, Universidad del Atl\'antico, Carrera 30 N\'umero 8-49\\
Puerto Colombia-Atl\'antico, Colombia} 

\date{}
\begin{document}
\maketitle

\begin{abstract}
\noindent In this work,  we study  inflation in a particular scalar-vector-tensor theory of gravitation without the $U(1)$ gauge symmetry.
The model is constructed from the more general action introduced in Heisenberg et al. (Phys Rev D 98:024038, 2018) using certain specific choices  for the Lagrangians and the coupling functions. Also, for this model  we build  the explicit form for the action, and from it, we derive the general equations: the energy-momentum tensor and the equations of motion, and  using the flat FLRW background, we have analyzed if it's possible  to obtain an inflationary regime with it. Additionally, using particular choices for the potential, the coupling functions, suitable dimensionless coupling constants  and  initial conditions, it was possible verify numerically that this model of inflation is viable. In this sense, we could verify that the introduction of the coupling function $f(\phi)$ in our model of inflation, allows us to reach a suitable amount of $e$-foldings $N$ for sufficient inflation. This is a remarkable result, since without the coupling function contribution,  the amount of $e$-foldings is smaller at the end of inflation, as has been demonstrated in Heisenberg et al. (2018). Also, the no-ghosts and stability conditions that the model during inflation must satisfy, i.e., absence of ghosts and Laplacian instabilities of linear cosmological perturbations were obtained, furthermore these conditions were verified numerically too.
\end{abstract}


\section{Introduction}
\noindent The  hypothesis of inflation is currently considered as a part of the standard modern cosmology. This hypothesis was introduced  in the early 1980's to resolve some problems that the Hot Big Bang model of the universe can't explain (e.g., the horizon, the flatness, and the monopole problems, among others) \cite{alexei1, guth, albrecht, linde}. Moreover, the inflationary paradigm allows us to make predictions about  properties of the current universe which are in excellent agreement with the cosmological and astrophysical  observations (e.g., the temperature fluctuations in CMB spectrum \cite{hu, soda, wands}, the existence of large scale structures \cite{chibisov, alexei2, hawking, guth2}, and the nearly scale invariant primordial power spectrum \cite{alexei3, lyth, lidsey}).\\
\noindent The origin or source of inflation is still now unknown. However, there are many  proposals in the literature  to explain the inflationary regime. For example, models with a scalar field (the so called inflaton field) minimally  coupled to gravity and with a nearly flat potential have been considered \cite{linde2, liddle1}.  In general, the inflaton field it can be considered  non-minimally coupled to gravity too (via the Ricci scalar, the Ricci tensor, the Gauss-Bonnet invariant, etc). In this sense,  the resulting models are known as scalar-tensor theories of gravitation.  Others proposals include Einstein–aether, and Bimetric theories, as well as TeVeS, $f(R)$, general higher-order theories, Ho\v{r}ava–Lifschitz gravity, Galileons, Ghost Condensates, and models of extra dimensions including  Kaluza-Klein, Randall–Sundrum and DGP (see e.g. Refs. \cite{odintsov1, clifton, odintsov2} for a review). In addition, in recent years a considerable amount of works have been considered to model the inflationary scenario using vector fields, e.g.  the generalized Proca fields and extensions thereof, or several Proca fields, as well as bigravity theories and scalar-vector-tensor (SVT) theories  (for more
details about these topics see Ref. \cite{lavinia1} and references therein). The SVT theories were introduced in Ref. \cite{lavinia2} and them constitute a consistent ghost-free covariant  gravity theories with second order equations of motion with derivative interactions. These SVT
theories have important implications for cosmological and astrophysical applications \cite{lavinia3, lavinia4, kase, lavinia5, ikeda}.\\
\noindent In this work we focused in study inflation in a particular SVT theory of gravitation without the $U(1)$ gauge symmetry.
This model is constructed from the more general action presented in Ref. \cite{lavinia4} using certain specific choices  for the Lagrangians and the coupling functions. Also, we derive the no-ghosts and stability conditions that the model during inflation must satisfy, i.e. absence of ghosts and Laplacian instabilities of linear cosmological perturbations.\\
\noindent This paper it is organized as follows: in section \ref{sec_model} we introduce the scalar-vector-tensor  model of inflation, and for this model  we construct  the explicit form for the action, from which, we derive the general equations: the energy-momentum tensor and the equations of motion.  In section \ref{sec_cosmo}, we write the general equations obtained in section \ref{sec_model} for a concrete background (the flat FLRW background), and in this way, we verify if it is possible  to obtain an inflationary regime with this model. Also, in this section, using concrete choices for the potential, the coupling functions, suitable dimensionless coupling constants  and  initial conditions, we verify numerically if this model of inflation is viable.  In section \ref{stability},  the no-ghosts and stability conditions that the model during inflation must satisfy, i.e., absence of ghosts and Laplacian instabilities of linear cosmological perturbations, are obtained, moreover these conditions  are verified numerically too. Finally, some conclusions are exposed in section \ref{conclus}. 
\section{The model}\label{sec_model}
\noindent In this section, we shall present in brief some basic features of the SVT theories, and also, we derive the general equations for the model under consideration.\\
\noindent The most general SVT model with broken $U(1)$ gauge symmetry which is ghost-free, with second order equations of motion and  with derivative interactions is given by the action \cite{lavinia4}
\begin{equation}\label{eq1}
S_{SVT}=\int{d^4x\,\sqrt{-g}\sum_{n=2}^{6}{\mathcal{L}_n}},
\end{equation}
where the Lagrangians $\mathcal{L}_2,\ldots,\mathcal{L}_6$, are given by
\begin{equation}\label{eq2}
\begin{aligned}
&\mathcal{L}_2=f_2(\phi,X_1,X_2,X_3,F,Y_1,Y_2,Y_3),\\
&\mathcal{L}_3=f_3(\phi,X_3)g^{\mu\nu}S_{\mu\nu}+\tilde{f}_3(\phi,X_3)A^{\mu}A^{\nu}S_{\mu\nu},\\
&\mathcal{L}_4=f_4(\phi,X_3)R+f_{4,X_3}(\phi,X_3)[(\nabla_{\mu}A^{\mu})^2-\nabla_{\mu}A_{\nu}\nabla^{\nu}A^{\mu}],\\
&\mathcal{L}_5=f_5(\phi,X_3)G^{\mu\nu}\nabla_{\mu}A_{\nu}-\frac{1}{6}f_{5,X_3}(\phi,X_3)
[(\nabla_{\mu}A^{\mu})^3-3\nabla_{\mu}A^{\mu}\nabla_{\rho}A_{\sigma}\nabla^{\sigma}A^{\rho}
+2\nabla_{\rho}A_{\sigma}\nabla^{\gamma}A^{\rho}\nabla^{\sigma}A_{\gamma}]\\
&+\mathcal{M}_5^{\mu\nu}\nabla_{\mu}\nabla_{\nu}\phi+\mathcal{N}_5^{\mu\nu}S_{\mu\nu},\\
&\mathcal{L}_6=f_6(\phi,X_1)L^{\mu\nu\alpha\beta}F_{\mu\nu}F_{\alpha\beta}+\mathcal{M}_6^{\mu\nu\alpha\beta}\nabla_{\mu}\nabla_{\alpha}\phi\nabla_{\nu}\nabla_{\beta}\phi
+\tilde{f}_6(\phi,X_3)L^{\mu\nu\alpha\beta}F_{\mu\nu}F_{\alpha\beta}+\mathcal{N}_6^{\mu\nu\alpha\beta}S_{\mu\alpha}S_{\nu\beta},
\end{aligned}
\end{equation}
This model was originally introduced in \cite{lavinia2} and its cosmological implications were studied in \cite{lavinia4}.
Moreover, in \cite{lavinia3} the authors using this model with gauge-invariant derivative scalar-vector interactions and a non-minimal coupling to gravity, to analyze the properties of black holes  on a static and spherically symmetric background. Also, in \cite{kase}
the cosmology of SVT theories with parity invariance was studied, paying particular attention to the application to dark energy. Finally, in
\cite{lavinia5} the gauge-ready formulation of cosmological perturbations on the flat Friedmann-Lema\v{i}tre-Robertson-Walker (FLRW)
background was performed (taking into account a matter perfect fluid).\\
In concrete, in this work we consider the following special case for the action given by Eq. (\ref{eq1})
\begin{equation}\label{eq3}
S=\int{d^4x\,\sqrt{-g}(\mathcal{L}_2+\mathcal{L}_4+\mathcal{L}_5)},
\end{equation}
where $\mathcal{L}_2$, $\mathcal{L}_4$ and $\mathcal{L}_5$ are construct considering the following  particular choices for $f_2$, $f_4$, and $f_5$ in Eq. (\ref{eq2}):
\begin{equation}\label{eq4}
\begin{aligned}
&f_2=F+X_1-V(\phi)+\beta_mMX_2+\beta_AM^2X_3,\\
&f_4=\frac{M_{pl}^2}{2}+\beta_{G}X_3,\\
&f_5=f(\phi),
\end{aligned}
\end{equation}
and  $X_1=-\frac{1}{2}\nabla_{\mu}\phi\nabla^{\mu}\phi$, $X_2=-\frac{1}{2}A^{\mu}\nabla_{\mu}\phi$, $X_3=-\frac{1}{2}A_{\mu}A^{\mu}$,
 $F=-\frac{1}{4}F_{\mu\nu}F^{\mu\nu}$, ($F_{\mu\nu}=\nabla_{\mu}A_{\nu}-\nabla_{\nu}A_{\mu}$), $V(\phi)$ is a scalar potential, $M$ is a constant having a
dimension of mass, $M_{\rm Pl}$ is  the reduced Planck mass, and $\beta_{m}$, $\beta_{A}$, $\beta_{G}$ are dimensionless coupling constants.
$f(\phi)$ is an arbitrary coupling function. The other quantities and coupling functions that appear in Eq. (\ref{eq2}) are defined in
\cite{lavinia4}, but in our case they are taken to be 0. A model of inflation using $f_2$ and $f_4$ was analyzed in \cite{lavinia4}.
The introduction of the coupling function $f_5=f(\phi)$ in our model induces a new mixing between the scalar field and the vector field via the Einstein tensor (in addition to the term $\beta_m M X_2$ of $f_2$). On the other hand, it could have significant implications onto the dynamics of
inflation.  For example, the amount of $e$-foldings $N$  obtained with the model introduced in \cite{lavinia4} is not sufficient in the end of inflation. In this way, 
with our model we will try to improve the amount of $e$-foldings $N$, among others calculations. This analysis and others we will carry out  in the
next sections.\\
Now, from the above, the  action given by Eq. (\ref{eq3}) takes the following explicit form:
\begin{equation}\label{eq5}
\begin{aligned}
S = &\int {\rm d}^4 x \sqrt{-g} \bigg[\frac{M_{\rm Pl}^2}{2} R - \frac{1}{4} F_{\mu \nu} F^{\mu \nu} - \frac{1}{2} \beta_A M^2 A_\mu A^\mu
- \frac{1}{2} \nabla_\mu \phi \, \nabla^\mu \phi - V(\phi) - \frac{1}{2} \beta_G A_\mu A^\mu R + \beta_G [(\nabla_{\mu} A^{\mu})^2\\
& -\nabla_{\mu} A_{\nu} \nabla^{\nu} A^{\mu}] - \frac{1}{2} \beta_m M A^\mu \nabla_\mu \phi + f(\phi) G_{\mu \nu} \nabla^\mu A^\nu \bigg],
\end{aligned}
\end{equation}
However, taking into account the identity
\begin{equation}\label{eq6}
\int {\rm d}^4 x \sqrt{-g} \Big[ (\nabla_{\mu} A^{\mu})^2 - \nabla_{\mu} A_{\nu} \nabla^{\nu} A^{\mu} \Big] = \int {\rm d}^4 x \sqrt{-g} R_{\mu \nu} A^\mu A^\nu,
\end{equation}
the action Eq. (\ref{eq5}) is reduced to
\begin{equation}\label{eq7}
\begin{aligned}
S = &\int {\rm d}^4 x \sqrt{-g} \bigg[ \frac{M_{\rm Pl}^2}{2} R - \frac{1}{2} \nabla_\mu \phi \, \nabla^\mu \phi - V(\phi)
- \frac{1}{4} F_{\mu \nu} F^{\mu \nu}- \frac{1}{2} \beta_A M^2 A_\mu A^\mu - \frac{1}{2} \beta_m M A^\mu \nabla_\mu \phi\\
&+ \Big( \beta_G  A^\mu A^\nu + f(\phi)  \nabla^\mu A^\nu \Big) G_{\mu \nu} \bigg],
\end{aligned}
\end{equation}
where $G_{\mu\nu}=R_{\mu\nu}-\frac{1}{2}R g_{\mu\nu}$ is the Einstein tensor. We can see in Eq. (\ref{eq7}) that the action $S$ can be written as $S=S_{HE}+S_{\phi}+S_{A}+S_{\phi A}$, where $S_{HE}$
stands the Hilbert–Einstein action, $S_{\phi}$ represents the pure scalar field sector of the action, $S_{A}$ includes the pure vector field sector and a term coupled
to the Einstein tensor. $S_{\phi A}$ denotes the sector which include the coupling between the scalar field $\phi$ and the vector field $A^{\mu}$ 
(via  derivative scalar-vector interactions) along with a term coupled to the Einstein tensor too. From this last sector, we will hope obtain new contributions to the background
equations that allow us study  the inflationary paradigm.\\
Since we have a explicit form for the action, Eq. (\ref{eq7}), then in this paper we will employ the usual algorithm to obtain the general  equations of motion, which is different to that used in \cite{lavinia4}, however, the results are equivalent.\\ The variation of the action Eq. (\ref{eq7}) with respect to the metric tensor $g_{\mu\nu}$ gives the field equations
\begin{equation}\label{eq8}
R_{\mu \nu} - \frac{1}{2} R g_{\mu \nu} = 8 \pi G T_{\mu \nu},
\end{equation}
where $T_{\mu\nu}$ is the energy–momentum tensor which is constructed from $S_{\phi}+S_{A}+S_{\phi A}$, and has the following form:
\begin{equation}\label{eq9}
T_{\mu \nu} = - \frac{2}{\sqrt{-g}} \frac{\delta S}{\delta g^{\mu \nu}} := T_{\mu \nu}^{(\phi)}+T_{\mu \nu}^{(A)}+ T_{\mu \nu}^{(A \phi)},
\end{equation}
Each term is given by the following expressions:
\begin{equation}\label{eq10}
 T_{\mu \nu}^{(\phi)} = \nabla_{\mu} \phi \, \nabla_{\nu} \phi - g_{\mu \nu} \left( \frac{1}{2} \nabla_\alpha \phi \, \nabla^\alpha \phi + V(\phi) \right),
\end{equation}
\begin{equation}\label{eq11}
\begin{aligned}
T_{\mu \nu}^{(A)}&=F_{\mu \alpha} F_{\nu}^{\,\,\, \alpha} + \beta_A M^2 A_\mu A_\nu + \beta_G \Big( R_{\mu \nu} A_\alpha A^\alpha- 4 G_{(\mu \alpha} A_{\nu )} A^\alpha
- R A_{\mu} A_{\nu} + 2 \nabla_{\alpha} \nabla_{(\mu} \left( A_{\nu)} A^\alpha  \right) - \Box \left( A_\mu A_\nu \right)\\
&- \nabla_{\mu} \nabla_{\nu} \left( A_\alpha A^\alpha \right) \Big)
 - g_{\mu \nu} \bigg[ \frac{1}{4} F_{\alpha \beta} F^{\alpha \beta}+ \frac{1}{2} \beta_A M^2 A_\alpha A^\alpha + \beta_G \Big( \nabla_{\alpha} \nabla_{\beta} \left( A^\beta A^\alpha \right)
- \Box \left( A_\alpha A^\alpha \right) - G_{\alpha \beta} A^\alpha A^\beta \Big) \bigg],
\end{aligned}
\end{equation}
\begin{equation}\label{eq12}
\begin{aligned}
 T_{\mu \nu}^{(A \phi)}& = \beta_m M A_{(\mu} \nabla_{\nu)} \phi + f(\phi) \Big[ R_{\mu \nu} \nabla_{\alpha} A^{\alpha} - R \, \nabla_{(\mu} A_{\nu)}
- 2 G_{(\mu \alpha} \nabla_{\nu)} A^\alpha \Big] + 2 G_{(\mu \alpha} A_{\nu)} \nabla^{\alpha} f(\phi)\\
& - \nabla_{\alpha} \big[ f(\phi) G_{\mu \nu} A^\alpha \big]
- \nabla_{\mu} \nabla_{\nu} \big[ f(\phi) \nabla_{\alpha} A^{\alpha} \big]+ \nabla_{\alpha} \nabla_{(\mu} \left[ f(\phi) \nabla^{\alpha} A_{\nu)} \right]
+ \nabla_{\alpha} \nabla_{(\mu} \left[ f(\phi) \nabla_{\nu)} A^{\alpha} \right]\\
& - \Box \left[ f(\phi) \nabla_{(\mu} A_{\nu)} \right]
 - g_{\mu \nu} \bigg[ \frac{1}{2} \beta_m M A^\alpha \nabla_\alpha \phi - f(\phi) G_{\alpha \beta} \nabla^\alpha A^\beta
+ \nabla_{\alpha} \nabla_{\beta} \left[ f(\phi) \nabla^{\beta} A^{\alpha} \right]
- \Box \left[ f(\phi) \nabla_{\alpha} A^{\alpha} \right] \bigg].
\end{aligned}
\end{equation}
On the other hand, the variation of the action with respect to $\phi$ gives the equation of motion,
\begin{equation}\label{eq13}
\Box \phi - V_{,\phi} + \frac{1}{2} \beta_m M \nabla_\mu A^\mu + G_{\mu \nu} \nabla^\mu A^\nu f_{,\phi} = 0,
\end{equation}
where $f_{,\phi}$ denotes derivative of $f$ with respect to $\phi$. In a similar way, the equation of motion for the vector field is given by
\begin{equation}\label{eq14}
\nabla^{\nu} F_{\mu \nu} + \frac{1}{2} \beta_m M \nabla_\mu \phi + \beta_A M^2 A_\mu - \Big( 2 \beta_G A^\nu - \nabla^\nu f(\phi) \Big) G_{\mu \nu} = 0.
\end{equation}
\section{Inflation in a SVT theory}\label{sec_cosmo}
\begin{figure*}
\centerline{\includegraphics[width=0.85\textwidth]{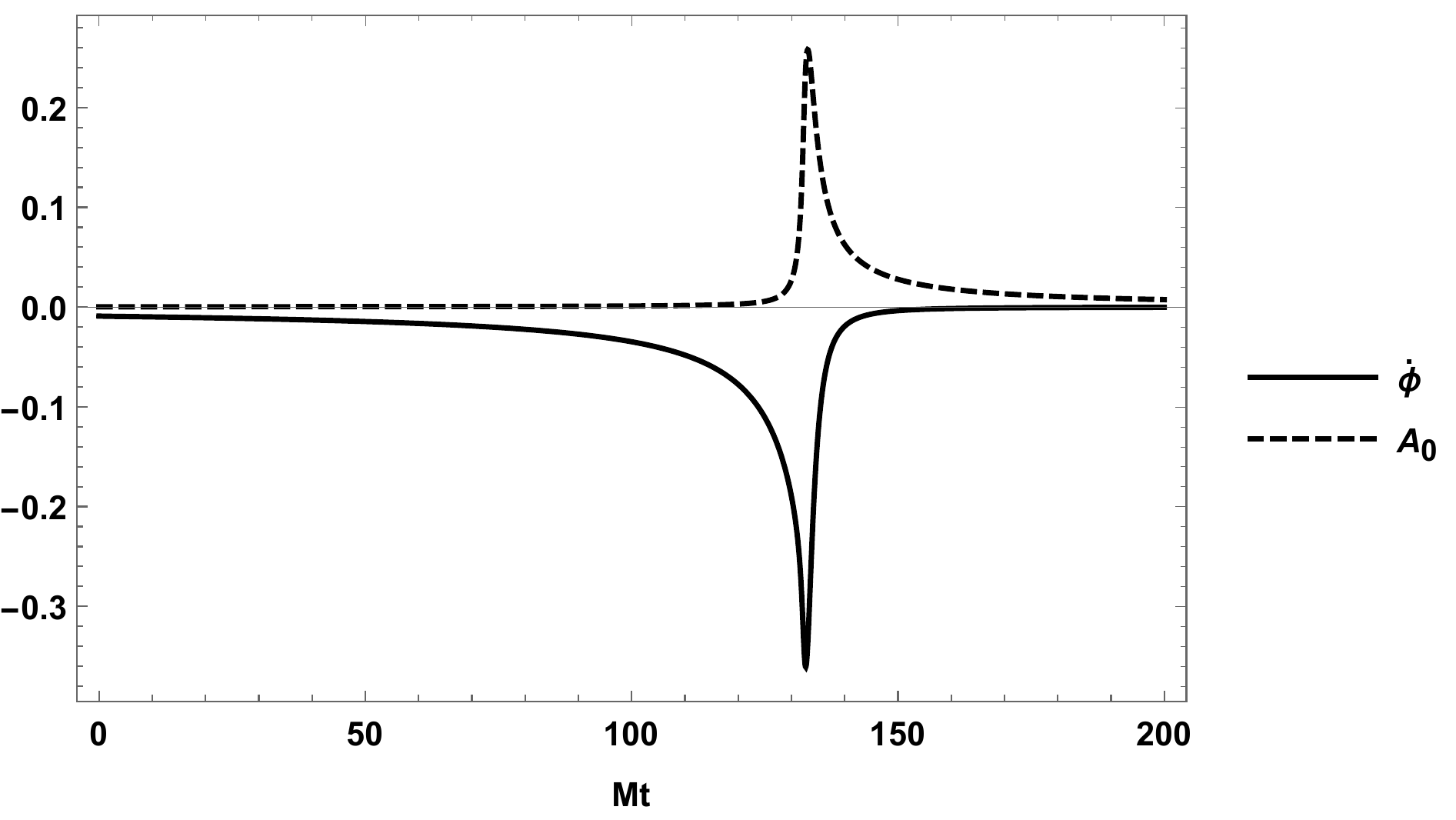}}
\caption{Evolution of $\dot{\phi}$  and $A_0$ (normalized by $M M_{\rm Pl}$ and $M_{\rm Pl}$, respectively) during  inflation and reheating. For this numerical simulation,  we have used the potential given by Eq. (\ref{eq30}) with $\alpha_c=\sqrt{6}/3$ (i.e., standard Starobinsky inflation) and the first coupling function given by Eq. (\ref{eq30'}) with $n=4$, $\omega=1\,M^4$. Moreover,  $\beta_m=0.08$, $\beta_A=1$, $\beta_G=0.3$ and the initial conditions $\phi(0)=5.5\,M_{\rm Pl}$, $\dot{\phi}(0)=-9\times 10^{-3}\,MM_{\rm Pl}$ and
$H(0)=0.494407\,M$, which is consistent with Eqs. (\ref{eq17}) and (\ref{eq21}).
\label{fig1}}
\end{figure*}

\begin{figure*}
\centerline{\includegraphics[width=0.85\textwidth]{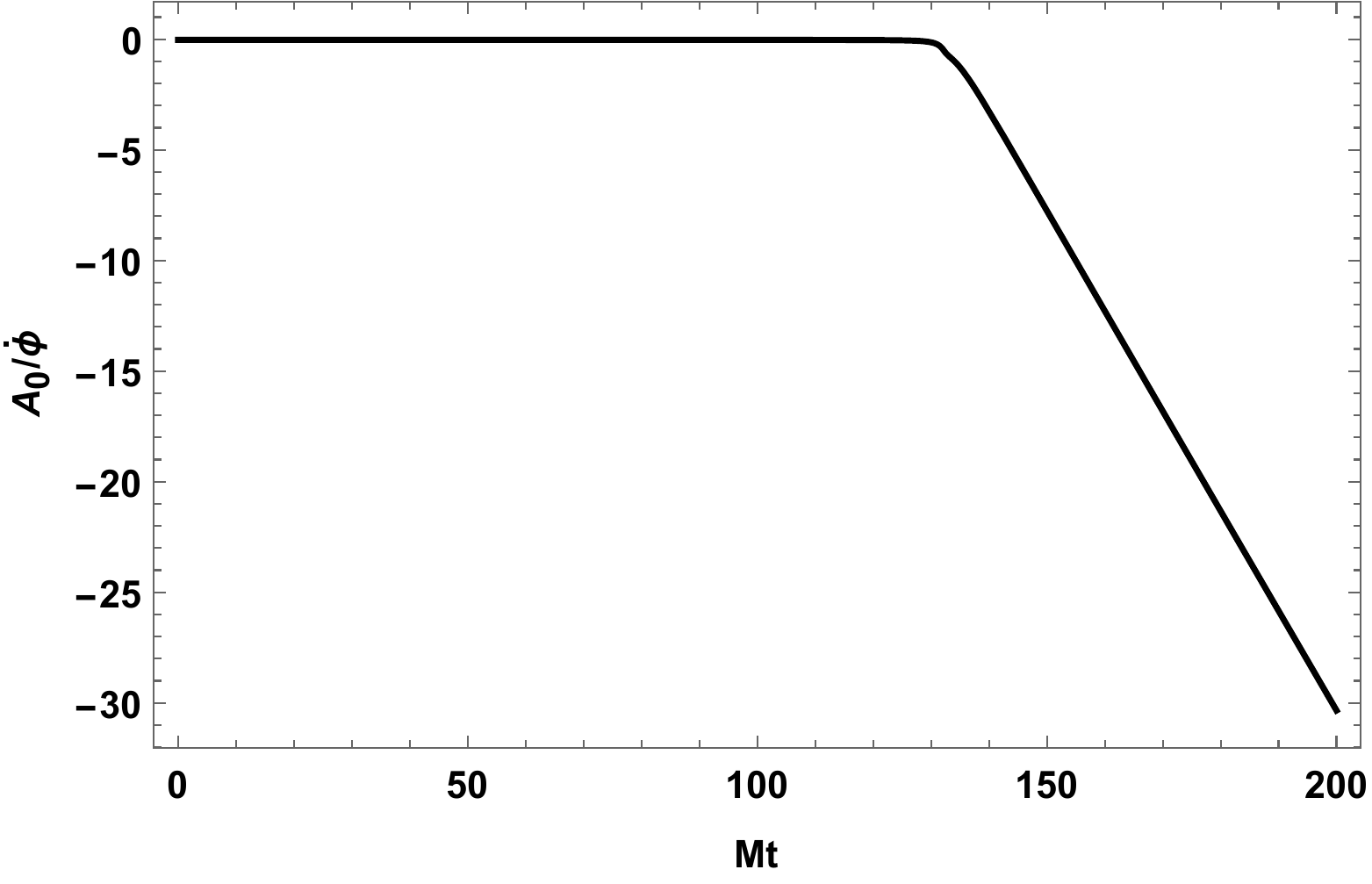}}
\caption{Evolution of $A_0/\dot{\phi}$ using the potential and the first coupling function given by Eqs. (\ref{eq30}) and  (\ref{eq30'}). In this numerical simulation the values of model parameters and initial conditions are the same as those used in Fig. \ref{fig1}.
\label{fig(2)}}
\end{figure*}

\begin{figure*}
\centerline{\includegraphics[width=0.85\textwidth]{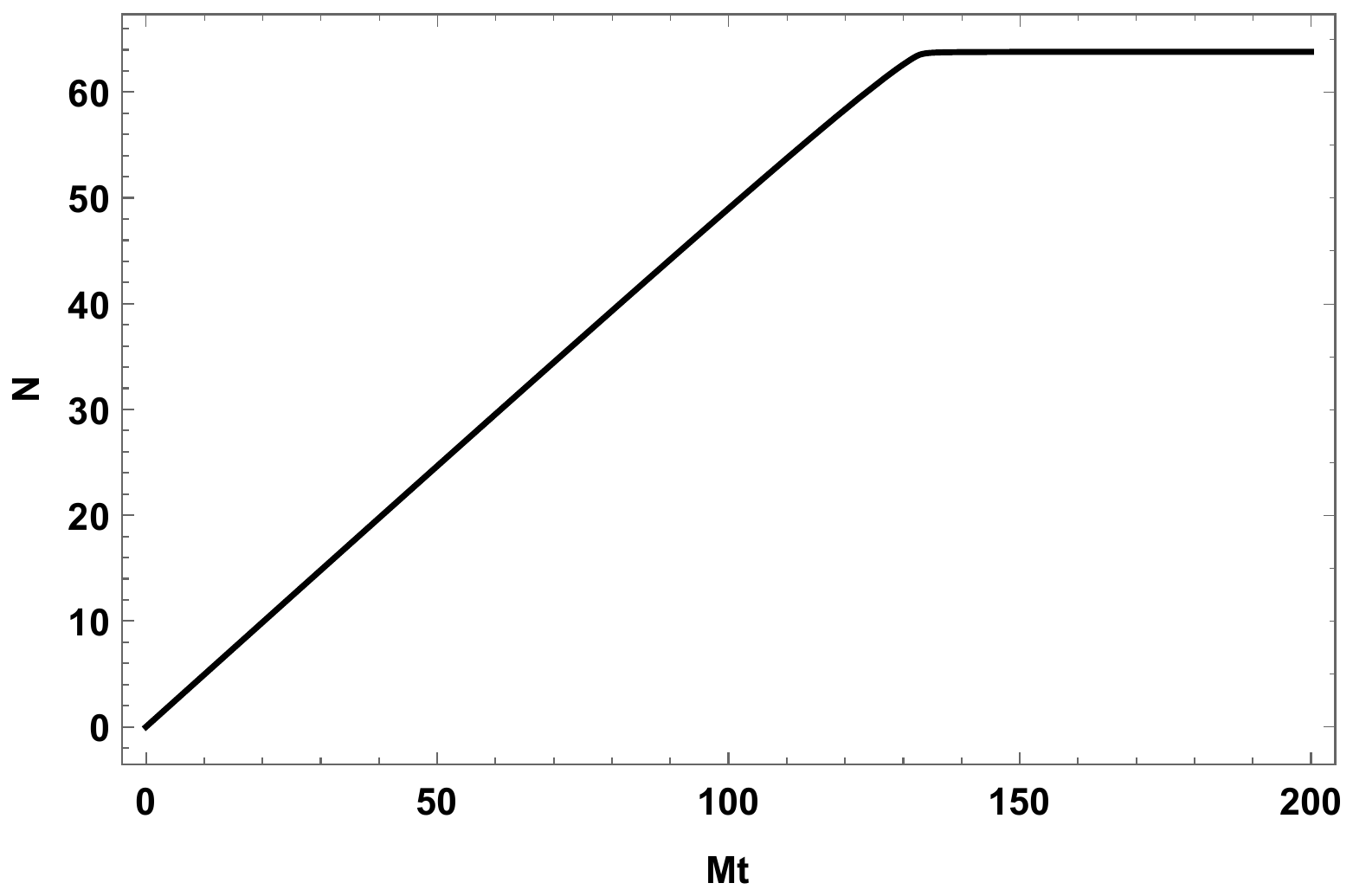}}
\caption{Evolution of the number of $e$-foldings $N$ using the potential and the first coupling function given by Eqs. (\ref{eq30}) and  (\ref{eq30'}). In this numerical simulation the values of model parameters and initial conditions are the same as those used in Fig. \ref{fig1}.
\label{fig2}}
\end{figure*}

\begin{figure*}
\centerline{\includegraphics[width=0.85\textwidth]{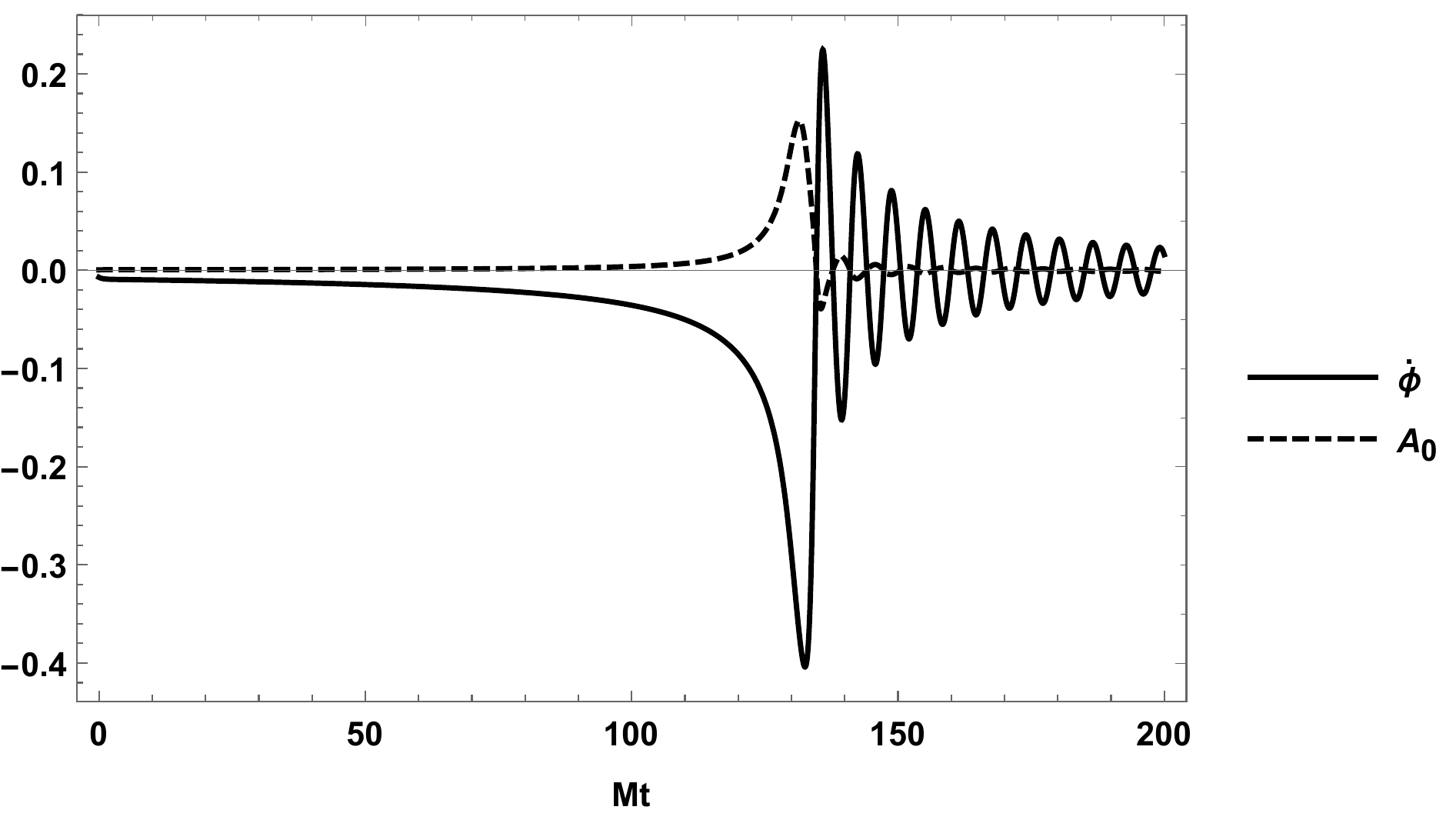}}
\caption{Evolution of $\dot{\phi}$  and $A_0$ (normalized by $M M_{\rm Pl}$ and $M_{\rm Pl}$, respectively) during  inflation and reheating. For this numerical simulation,  we have used the potential given by Eq. (\ref{eq30}) with $\alpha_c=\sqrt{6}/3$ (i.e., standard Starobinsky inflation) and the second coupling function given by Eq. (\ref{eq30'}) with $\xi=4.5$, $\lambda=1$. Additionally, $\beta_m=0.08$, $\beta_A=1$, $\beta_G=0.01$ and the initial conditions $\phi(0)=5.5\,M_{\rm Pl}$, $\dot{\phi}(0)=-9\times 10^{-3}\,MM_{\rm Pl}$ and
$H(0)=0.132553\,M$, which is consistent with Eqs. (\ref{eq17}) and (\ref{eq21}).
\label{fig3}}
\end{figure*}

\begin{figure*}
\centerline{\includegraphics[width=0.85\textwidth]{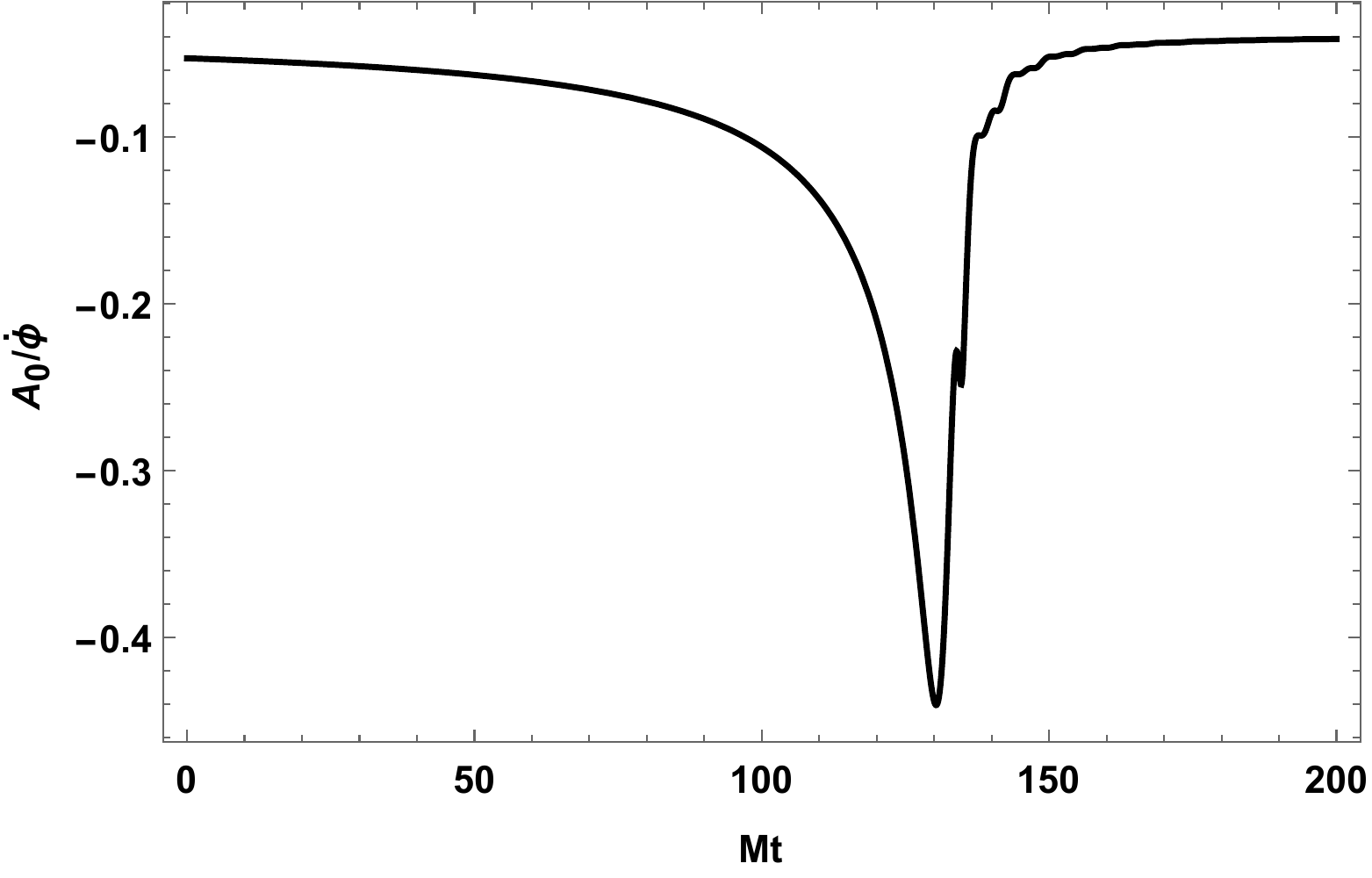}}
\caption{Evolution of $A_0/\dot{\phi}$ for the potential and the second coupling function given by Eqs. (\ref{eq30}) and (\ref{eq30'}). In this numerical simulation the values of model parameters and initial conditions are the same as those used in Fig. \ref{fig3}.
\label{fig(4)}}
\end{figure*}

\begin{figure*}
\centerline{\includegraphics[width=0.85\textwidth]{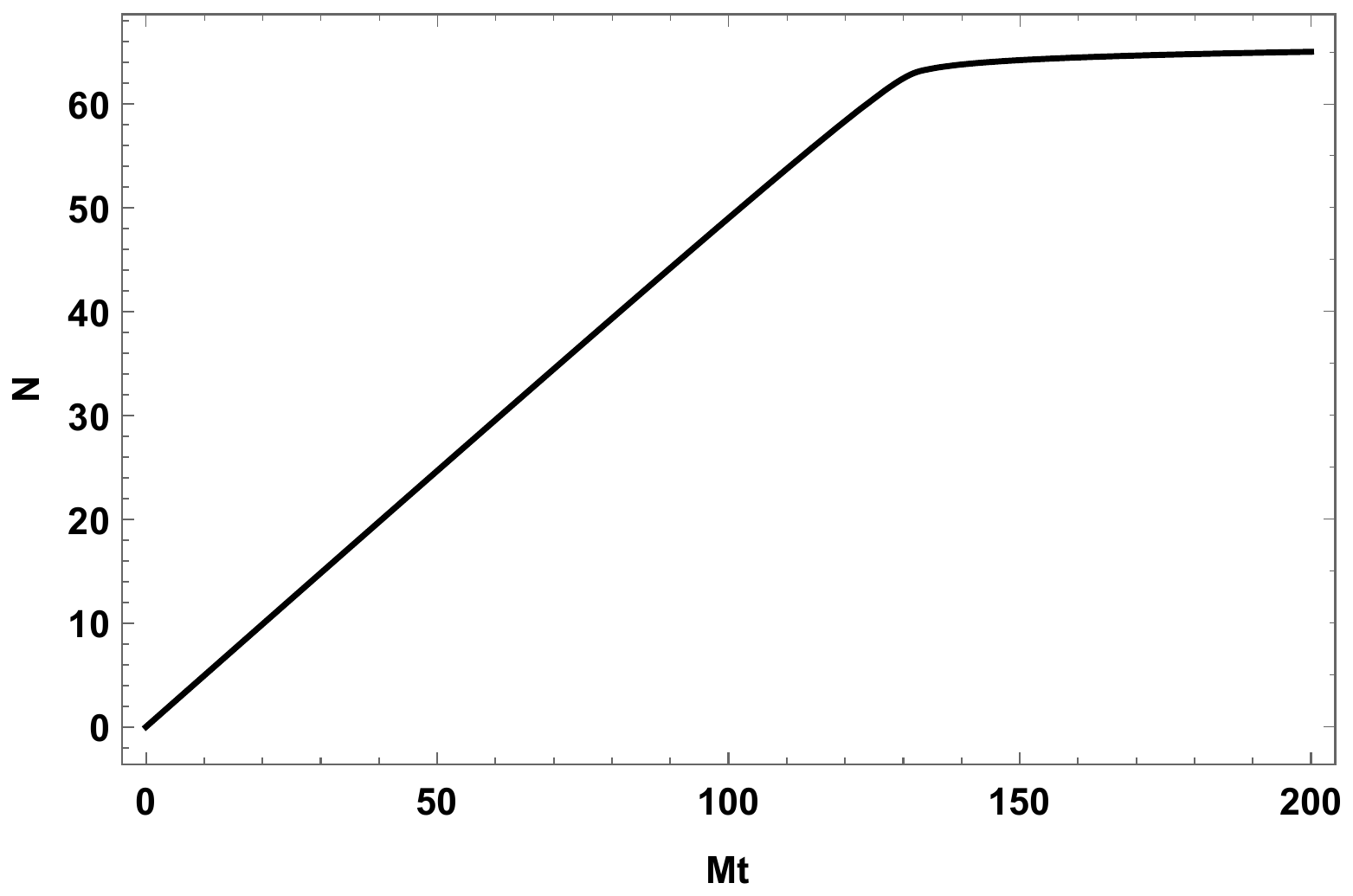}}
\caption{Evolution of the number of $e$-foldings $N$ for the potential and the second coupling function given by Eqs. (\ref{eq30}) and (\ref{eq30'}). In this numerical simulation the values of model parameters and initial conditions are the same as those used in Fig. \ref{fig3}.
\label{fig4}}
\end{figure*}

\begin{figure*}
\centerline{\includegraphics[width=0.85\textwidth]{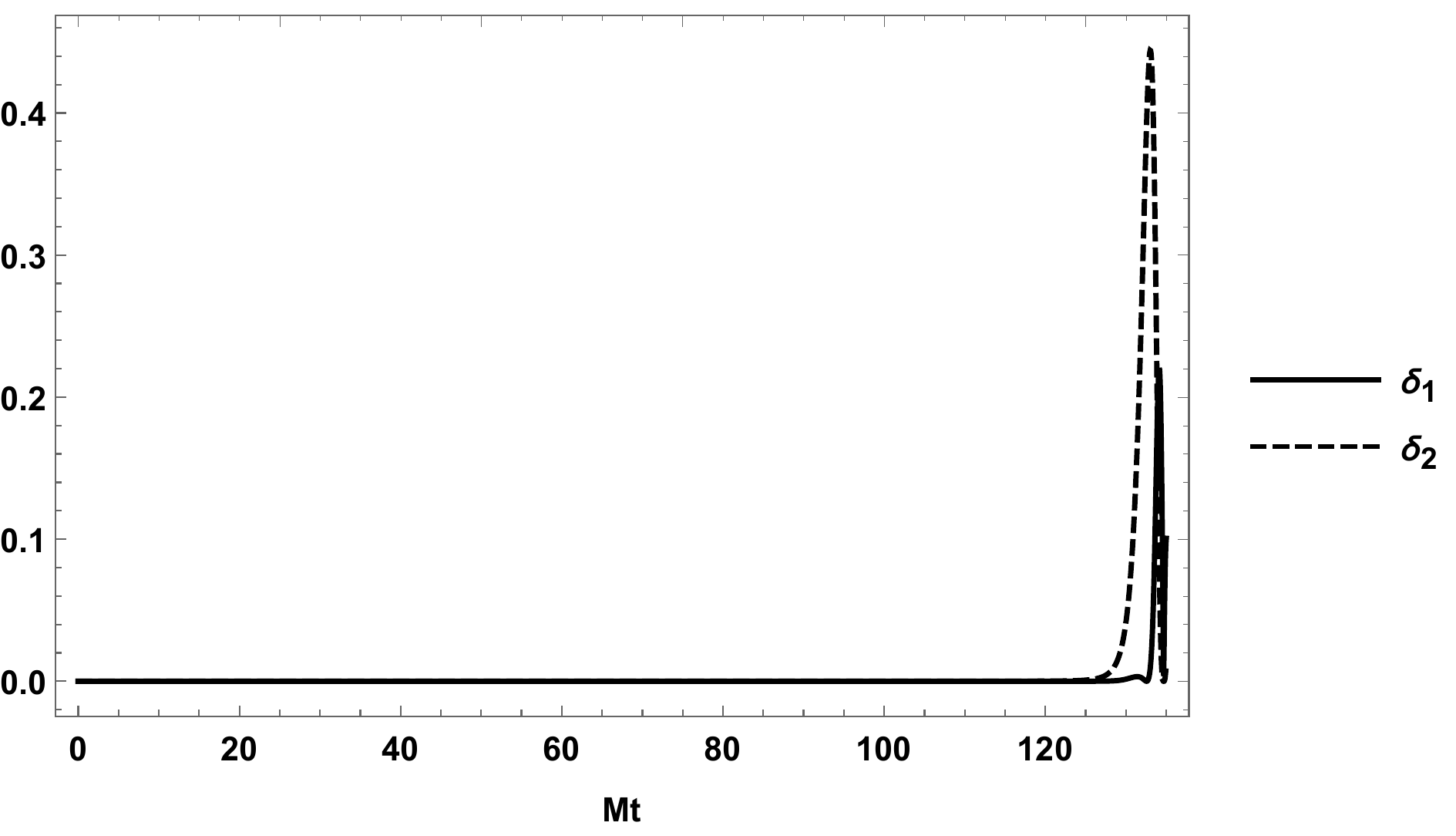}}
\caption{Evolution of slow-roll parameters $\delta_1$ and $\delta_2$ during  inflation. In this numerical simulation we have used the potential, the coupling function, the values of model parameters and initial conditions as those used in Fig. \ref{fig3}.
\label{fig5}}
\end{figure*}

\begin{figure*}
\centerline{\includegraphics[width=0.85\textwidth]{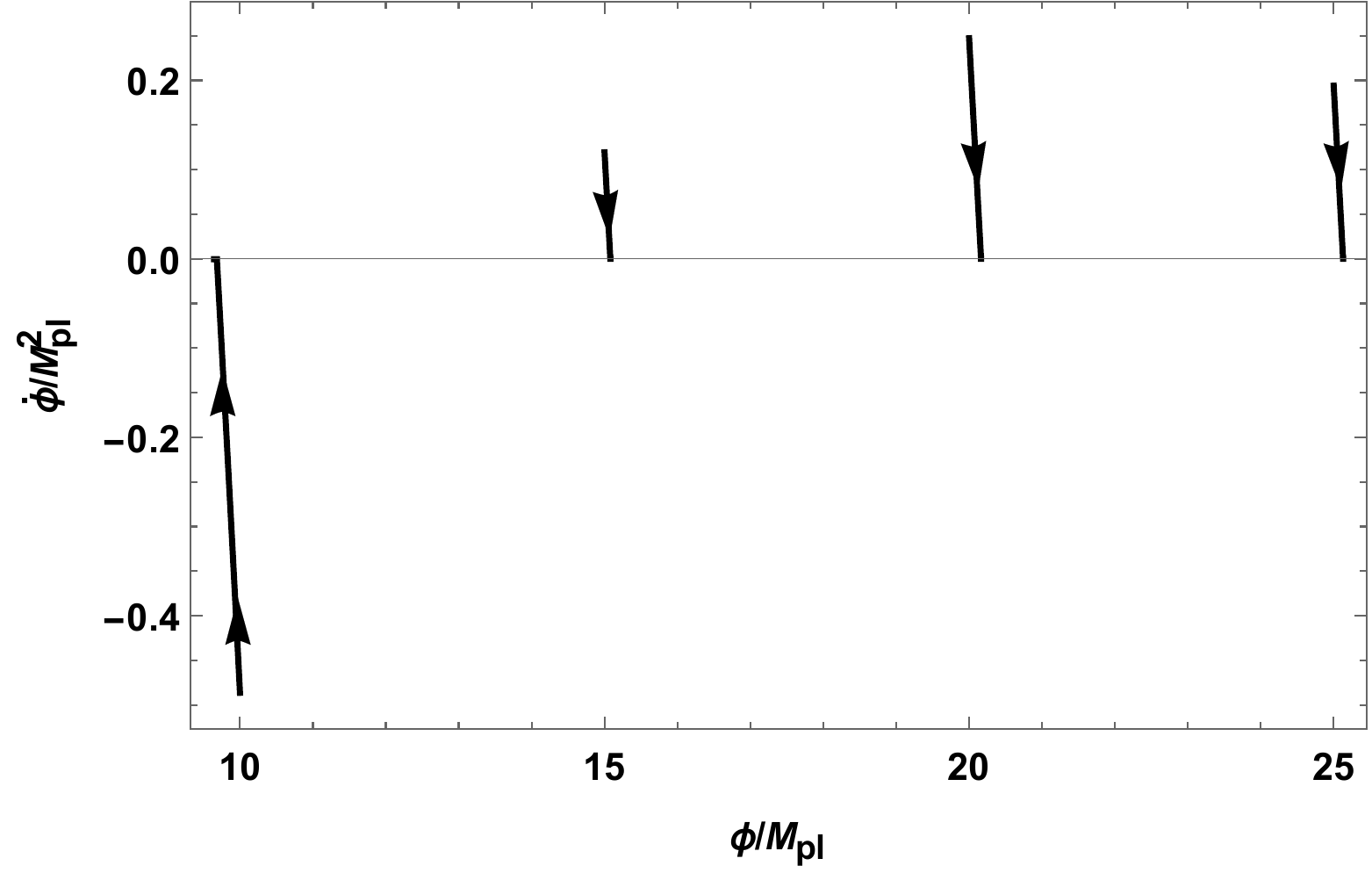}}
\caption{Phase-space structure for the inflationary solution. In this numerical simulation we have used the potential, the coupling function, the values of model parameters as those used in Fig. \ref{fig3},  and various initial conditions for $\phi$ were used.\label{fig6}}
\end{figure*}

\noindent In this section we use the general equations obtained above in a concrete background and in this way we will see if possible  
to obtain a inflationary regime with this model. The usual choice for the background is given by the flat Friedmann-Robertson-Walker (FRW) metric whose line element has the following form
\begin{equation}\label{eq15}
ds^2 = - dt^2 + a^2(t) \delta_{ij} dx^i dx^j,
\end{equation}
where $a(t)$ is the scale factor. In order to guarantee the homogeneity and isotropy of the  universe,  in the next calculations, we will regard that the vector field is without spatial components, i.e.
\begin{equation}\label{eq16}
A_\mu(t) =( A_0 (t) , 0, 0, 0), 
\end{equation}  
and the scalar field is time dependent only, $\phi=\phi(t)$. Thereby, using the FRW metric and the vector and scalar fields defined previously,  the Einstein's equations Eq.  (\ref{eq8}) and the equations of motion Eqs. (\ref{eq13}) and (\ref{eq14}) are reduced to
\begin{equation}\label{eq17}
\begin{aligned}
3 M_{\rm Pl}^2 H^2 &=\frac{1}{2} \dot{\phi}^2 + V(\phi) + \frac{1}{2} \beta_A M^2 A_0^2 + 9 \beta_G H^2 A_0^2
+ \frac{1}{2}\beta_m M A_0 \dot{\phi} - 9 H^2 A_0 \dot{\phi}f_{,\phi},
\end{aligned}
\end{equation}

\begin{equation}\label{eq18}
\begin{aligned}
- \big( 2 \dot{H} + 3  H^2 \big) M_{\rm Pl}^2 &= \frac{1}{2} \dot{\phi}^2 - V(\phi) + \frac{1}{2} \beta_A M^2 A_0^2 - 3 \beta_G H^2 A_0^2
- 4 \beta_G H A_0 \dot{A}_0 - 2 \beta_G \dot{H} A_0^2 + \frac{1}{2} \beta_m M A_0 \dot{\phi}\\
&+ 3 H^2 A_0 \dot{\phi}f_{,\phi}+ 2 H \dot{A}_0 \dot{\phi} f_{,\phi}+ 2 \dot{H} A_0 \dot{\phi} f_{,\phi}
+ 2 H A_0 \ddot{\phi} f_{,\phi}+ 2 H A_0 \dot{\phi}^2 f_{,\phi\phi},
\end{aligned}
\end{equation}
\begin{equation}\label{eq19}
\begin{aligned}
&\ddot{\phi} + 3 H \dot{\phi} + V_{,\phi} + \frac{1}{2} \beta_m M \Big( \dot{A}_0 + 3 H A_0 \Big)
- 3 H \Big( 3 H^2 A_0 + H \dot{A}_0 + 2 \dot{H} A_0 \Big)f_{,\phi} = 0,
\end{aligned}
\end{equation}
\begin{equation}\label{eq20}
\frac{1}{2} \beta_m M \dot{\phi} + \beta_A M^2 A_0 + 3 H^2 \left( 2 \beta_G A_0 - \dot{\phi}f_{,\phi} \right) = 0.
\end{equation}
It is clear that the introduction of the coupling function $f(\phi)$ in our model gives rise to new and important terms in the background equations (although they are more complicated). Additionally, making suitable choices for  $V(\phi)$, $f(\phi)$, the dimensionless coupling
constants $\beta_{m}$, $\beta_{A}$, $\beta_{G}$ and appropriated initial conditions, is possible in principle, to obtain a viable inflationary regime with this model.\\
Now,  from Eq. (\ref{eq20}), we can to write the temporal vector component $A_0$ as
\begin{equation}\label{eq21}
A_0=\frac{-\beta_m M+6H^2f_{,\phi}}{2(\beta_A M^2+6\beta_G H^2)}\dot{\phi},
\end{equation}
Is evident from Eq. (\ref{eq21}) that there is a direct explicit relation between the temporal vector component $A_0$ and $\dot{\phi}$, and
an implicit relation with $\phi$ (through of the coupling function $f(\phi)$). Therefore, in this case we can say that the dynamics of inflation is driven by the scalar field $\phi$ and its derivative $\dot{\phi}$. If we consider that
\begin{equation}\label{eq22}
f_{,\phi}=\frac{\partial f}{\partial \phi}=\frac{\beta_m M}{6H^2}, 
\end{equation}
then $A_0=0$ and the model corresponds to that with a single scalar field (inflaton) minimally coupled to gravity, i.e. the action
Eq. (\ref{eq7}) it reduces to $S_{HE}+S_{\phi}$, which has been widely studied in the literature. Now, considering a pure de Sitter expansion for the inflationary stage,  we can found the coupling function $f(\phi)$ integrating  Eq. (\ref{eq22}) (since  $H$ is constant in this case) 
\begin{equation}\label{eq23}
f(\phi)=\frac{\beta_m M}{6H^2}\phi+C, 
\end{equation}
where $C$ is an integration constant.\\
\noindent In order to analyze the inflationary regime with the background Eqs. (\ref{eq17})-(\ref{eq20}),
we consider the  slow-roll parameters defined as
\begin{equation}\label{eq24}
\epsilon\equiv -\frac{\dot{H}}{H^2},\ \ \ \ \epsilon_V\equiv\frac{M_{\rm Pl}^2}{2}\left(\frac{V_{,\phi}}{V}\right)^2,\ \ \ \ \eta\equiv\frac{\ddot{\phi}}{H\dot{\phi}},
\end{equation}
additionally,  during the inflationary phase these slow-roll parameters must satisfy the restriction $|\epsilon|\ll 1$, $|\epsilon_V|\ll 1$
and $|\eta|\ll 1$ \cite{wands}. Besides, in our case we define the following slow-roll parameters
\begin{equation}\label{eq25}
|\delta_1|\equiv\left|\frac{A_0\ddot{\phi}f_{,\phi}}{H}\right|\ll 1, \ \ \ \ |\delta_2|\equiv\left|\frac{A_0\dot{\phi}^2f_{,\phi\phi}}{H}\right|\ll 1.
\end{equation}
Replacing Eq. (\ref{eq21}) and its time derivative into Eq. (\ref{eq18}) and using Eq. (\ref{eq17}) to eliminate $V$, 
the slow-roll parameter $\epsilon$ takes the form
\begin{equation}\label{eq26}
\begin{aligned}
\epsilon=&\Big\{\dot{\phi}^2(\beta_A M^2+6\beta_G H^2)\Big[\beta_A M^4(4\beta_A-\beta_m^2)
+2H\Big(M^2(24\beta_A\beta_G-\beta_G\beta_m^2(3+2\eta)
+2M\beta_A\beta_m(3-2\eta)f_{,\phi}H\\
&+6(12\beta_G^2+Mf_{,\phi}(6\beta_G\beta_m+\beta_A M(4\eta-3)f_{,\phi})H^3
+36\beta_G(2\eta-3)f_{,\phi}^2H^5+2(-\beta_A\beta_mM^3
+12f_{,\phi}(M^2\beta_AH^2\\
&+3\beta_GH^4))f_{,\phi\phi}\dot{\phi}\Big)\Big]\Big\}/
\Big\{H^2\Big[8M_{\rm Pl}^2(\beta_AM^2+6\beta_GH^2)^3
+2\Big(-M^4\beta_A\beta_m(\beta_G\beta_m+2M\beta_Af_{,\phi})+
18M^2(\beta_G^2\beta_m^2\\
&+2M\beta_Af_{,\phi}(\beta_G\beta_m
+M\beta_Af_{,\phi}))H^2+108M^2\beta_A\beta_Gf_{,\phi}^2H^4
+216\beta_G^2f_{,\phi}^2H^6\Big)\dot{\phi}^2\Big]\Big\},
\end{aligned}
\end{equation}
where we have used the slow-roll parameter $\eta$ to eliminate $\ddot{\phi}$, and $f_{,\phi\phi}$ depicts the second derivative of $f$ with
respect to $\phi$.\\
\noindent  Now, considering in Eqs. (\ref{eq19}) and (\ref{eq21}) the slow-low parameters given by Eqs. (\ref{eq24}) and (\ref{eq25}) in an inflationary regime,  then we can obtain an approximate expression for $\dot{\phi}$
\begin{equation}\label{eq27}
\dot{\phi}\approx \frac{4V_{,\phi}(M^2\beta_A+6\beta_GH^2)}
{3H\left[M^2(\beta_m^2-4\beta_A)-12(2\beta_G+M\beta_mf_{,\phi})H^2+36f_{,\phi}^2H^4\right]},
\end{equation}
For $f(\phi)=0$ the Eqs. (\ref{eq26}) and (\ref{eq27}) it reduces to those obtained in \cite{lavinia4}. For suitable values of the dimensionless coupling constants $\beta_{m}$, $\beta_{A}$, $\beta_{G}$ and some specific choices for the
coupling function $f(\phi)$, we can assume that the scalar potential $V$ dominates over the other terms on the right-hand side
of Eq. (\ref{eq17}) during slow-roll inflation. This assumption has sense, since all terms on the right-hand side
of Eq. (\ref{eq17}) (apart of $V$) are proportional to $\dot{\phi}^2$ (after of replacing $A_0$ by Eq. (\ref{eq21})) and as is well known, 
in the inflationary regime $\dot{\phi}^2\ll V$. In this way the Eq. (\ref{eq17}) it's reduces to
\begin{equation}\label{eq28}
3 M_{\rm Pl}^2 H^2\approx V(\phi),
\end{equation}
Replacing the last equation and Eq. (\ref{eq27}) into Eq. (\ref{eq26}), and neglecting the slow-roll parameters $\eta$ and
$\epsilon_V$ relative to 1 in the end. Thereby we get
\begin{equation}\label{eq29}
\epsilon\approx\epsilon_V\left(1+\frac{\beta_m^2M^2M_{\rm Pl}^4-4MM_{\rm Pl}^2\beta_mf_{,\phi}V+4f_{,\phi}^2V^2}{4\beta_AM^2M_{\rm Pl}^4+4M_{\rm Pl}^2(2\beta_G+\beta_m Mf_{,\phi})V-4f_{,\phi}^2V^2}\right),
\end{equation}
where in the last step, we have considered that $|\beta_m|\ll 1$. Additionally, making $\beta_m=0$ and $f(\phi)=0$ in Eq. (\ref{eq29}),
we have $\epsilon\approx\epsilon_V$ which corresponds to standard slow-roll inflation. We see that, the insertion of a mixing between 
$\dot{\phi}$ and $A_0$ in our model could have important consequences in the evolution of the inflationary parameters, e.g., 
$\dot{\phi}$, $A_0$, the number of $e$-foldings $N$, among others. \\
In order to verify if the above considerations are correct,  we will solve numerically the general background equations obtained before.
To do this, we replace Eq. (\ref{eq21}) (and its temporal derivative)  in Eqs. (\ref{eq17}) and (\ref{eq19}), and in this way the Eqs. (\ref{eq17}) and (\ref{eq19}) are in terms  of the scalar field $\phi$ and $H$. Moreover,  taking into account appropriated choices for the scalar potential $V(\phi)$ and the coupling function $f(\phi)$, for example
\begin{equation}\label{eq30}
V(\phi)=\frac{M^2M_{\rm Pl}^2}{2\alpha_c^2}\left(1-e^{-\alpha_c\phi/M_{\rm Pl}}\right)^2,
\end{equation}
where $\alpha_c$ is a positive constant. This potential corresponds to the so-called $\alpha$-attractor model \cite{linde3}. And, for the coupling function $f(\phi)$, in this work  we will consider two choices:
\begin{equation}\label{eq30'}
f(\phi)=\frac{\omega}{\phi^n}\ \ \ \text{and}\ \ \ f(\phi)=\xi e^{-\lambda\phi/M_{\rm Pl}},
\end{equation}
where in first coupling function, $\omega$ is an arbitrary parameter with dimension of $M^{n}$ ($n\in\mathbb{Z}$) and it is  type power-law coupling function, which has been widely studied in the literature (see \cite{granda}). In the second one,  $\xi$ 
and $\lambda$ are  arbitrary dimensionless parameters. This type of exponential coupling function has been used in the literature too (see \cite{odintsov3} and
\cite{fronimos}). Using the expressions given by Eq. (\ref{eq30}) and the first coupling function in Eq. (\ref{eq30'}), appropriated choices for the dimensionless coupling constants $\beta_{m}$, $\beta_{A}$, $\beta_{G}$ and along with appropriated initial conditions, we can check numerically if the proposed model is viable to describes  inflation.\\
\noindent In Fig. \ref{fig1} we can see the evolution of $\dot{\phi}$  and $A_0$  during inflation and reheating, and from it, is clear that $\dot{\phi}$  and $A_0$ display a slow evolution during inflation, which is a fundamental 
condition that must fulfill the fields in the inflationary regime. Also, we can see in Fig. \ref{fig(2)} that the ratio $A_0/\dot{\phi}$ obtained
from Eq. (\ref{eq21}), is nearly constant during inflation (i.e. for $Mt\lesssim 130$), since $H$ and $f_{,\phi}$ are nearly constant during this stage. Moreover, in Fig. \ref{fig1} we can see that the fields decay faster during reheating, but the ratio $A_0/\dot{\phi}$ is not nearly constant (see Fig. \ref{fig(2)} ).\\
\noindent  In Fig. \ref{fig2} we have plotted the evolution of the number of $e$-foldings $N$ and
we see that for appropriate values of the model parameters and initial conditions used here, it is possible to obtain a suitable amount of $e$-foldings ($N\gtrsim 60$) for sufficient inflation. In this case, $N\approx 63$ (for $\omega=1\,M^4$) at the end of inflation. In general, inflation end   when the slow-roll parameters  leave of be much smaller than 1 and in our case this happens from $Mt\approx 130$  (see Fig. \ref{fig5}).\\
\noindent  Using the second coupling function, in Fig. \ref{fig3}, we can see  a similar behavior during inflation to the shown in Fig. \ref{fig1}, but during reheating the fields oscillate and the amplitude of $A_0$ decreases faster than the amplitude of  $\dot{\phi}$. Furthermore, we can see in Fig. \ref{fig(4)} that the ratio $A_0/\dot{\phi}$ during inflation is not nearly constant, but its variation is small in this phase. Additionally, in Fig. \ref{fig4} is clear
that for this second choice of the coupling function, we get a higher amount of $e$-foldings  than  obtained 
using the first coupling function. In this case, $N\approx 64$ (for $\xi=4.5$) at the end of inflation. These results it can explain taking into account  that in both cases $A_0$ decay faster than $\dot{\phi}$ during inflation  (see Figs. \ref{fig1} and \ref{fig3}). Therefore, from the above numerical analysis,  we can say  that the introduction of the coupling function $f(\phi)$ in our model of inflation, allows us to reach a suitable amount of $e$-foldings $N$ for sufficient inflation. This result in very important, since without the coupling function contribution,  the amount of $e$-foldings is
smaller at the end of inflation. This result can be seen in Fig. 1 of \cite{lavinia4}. Finally, in Fig. \ref{fig5} we plotted the evolution of
the new slow-roll parameters $\delta_1$ and $\delta_2$ which are defined by Eq. (\ref{eq25}),  is clear that this parameters satisfy the conditions $|\delta_1|\ll 1$ and $|\delta_2|\ll 1$ during  inflation. If we consider the first coupling function given by Eq. (\ref{eq30'}),
a similar behavior is observed.\\
Now, in order to verify whether the numerical inflationary solution obtained above  is an attractor solution or not, we carry out an analysis on the phase-space. In this ways, in Fig \ref{fig6} we show the phase-space diagram with various initial conditions and we see that the phase-space flow converge in distinct points depending on initial conditions. This observed pattern is related to the form of the potential, since it is approximately constant during inflation (flat potential)  and it preserves its form under the translation $\phi \rightarrow \phi+\text{const}$. Therefore, we can say that the solution used here is independent of initial conditions (i.e. an attractor solution).

\section{No-ghosts and stability conditions}\label{stability}
\noindent In this section, we exhibit the no-ghosts and stability conditions that our model during inflation must satisfy, i.e., absence of ghosts and Laplacian instabilities of linear cosmological perturbations. Since the general expressions associated to this analysis are very complicated, in our case, this conditions will be verified numerically taking into account particular choices for the model parameters, the potential $V(\phi)$, the coupling functions $f(\phi)$ and the  initial conditions used in the previous section. For this task, we will use the general expressions for tensor, vector and scalar perturbations obtained in \cite{lavinia4}. In this way and in order to have a self-contained document, in the next sections we reproduce literally the more important results associated with the general perturbation analysis which was performed in the sections \textbf{III}, \textbf{IV} and \textbf{V} of \cite{lavinia4}, and then we apply this general formalism to the present model.
\subsection{Tensor perturbations} To start this analysis, it is consider the linearly perturbed line element of intrinsic tensor modes, which is given by
\begin{equation}\label{eqp1}
ds_t^2=-dt^2+a^2(t)(\delta_{ij}+h_{ij})dx^idx^j,
\end{equation}
where the tensor perturbation $h_{ij}$ obeys the usual transverse and traceless conditions $\nabla^jh_{ij}=0$ and $h^i_i=0$. Expanding the action (\ref{eq1}) up to quadratic order in $h_{ij}$ and integrating it by parts, the second-order action of tensor perturbations yields
\begin{equation}\label{eqp2}
S^{(2)}_t=\int{dtd^3x\frac{a^3q_t}{8}\delta^{ik}\delta^{jl}\left[\dot{h}_{ij}\dot{h}_{kl}
-\frac{c_t^2}{a^2}(\partial h_{ij})(\partial h_{kl})\right]},
\end{equation}
where the symbol $\partial$ represents the spatial partial derivative and  $q_t$ and $c_t^2$ are given by Eqs. (3.3) and (3.4) of \cite{lavinia4}. For the model under consideration, the associated quantities for tensor perturbations $q_t$ and $c_t^2$,  are given by
\begin{equation}\label{eq31}
\begin{aligned}
q_t=&M_{\rm Pl}^2-\frac{(\beta_m M-6f_{,\phi}H^2)(M\beta_G\beta_m+2f_{,\phi}(\beta_A M^2
+3\beta_G H^2))\dot{\phi}^2}{4(\beta_A M^2+6\beta_G H^2)^2},
\end{aligned}
\end{equation}
\begin{equation}\label{eq32}
\begin{aligned}
&c_t^2=1+\frac{2(\beta_m M-6f_{,\phi}H^2)(M\beta_G\beta_m
+2f_{,\phi}(\beta_A M^2+3\beta_GH^2))\dot{\phi}^2}{4M_{\rm Pl}^2(\beta_A M^2+6\beta_G H^2)^2
-(\beta_m M-6f_{,\phi}H^2)(M\beta_G\beta_m+2f_{,\phi}(\beta_A M^2+3\beta_G H^2))\dot{\phi}^2},
\end{aligned}
\end{equation}
where $c_t$ represent the propagation speed of gravitational waves on the FLRW background and  Eq. (\ref{eq21}) was used to 
express $A_0$ in terms of $\dot{\phi}$. Furthermore, to guarantee the lack of ghost or Laplacian instabilities in the tensor sector, it must be fulfilled that $q_t>0$ and $c_t^2>0$.\\
\subsection{Vector perturbations} For perturbations in the vector sector, it's used the perturbed line element in the flat gauge: 
\begin{equation}\label{eqp3}
ds^2_v=-dt^2+2V_idtdx^i+a^2(t)\delta_{ij}dx^idx^j,
\end{equation}
where $V_i$ is the vector perturbation obeying the transverse condition $\nabla^iV_i=0$. The temporal and spatial components of $A^{\mu}$ associated with the intrinsic vector sector are expressed as
\begin{equation}\label{eqp4}
A_0=A_0(t),\ \ \ \ \ A_i=Z_i(t,x^i),
\end{equation}
where $Z_i$ is the intrinsic vector perturbation which satisfies the transverse condition $\partial^iZ_i=0$. Expanding Eq. (\ref{eq1}) up to quadratic order in perturbations and using the background Eqs.
(2.13) and (2.16) of \cite{lavinia4}, the resulting second-order action in the vector sector yields
\begin{equation}\label{eqp5}
\begin{aligned}
S^{(2)}_v=&\int dtd^3x\sum_{i=1}^{2}\Big[\frac{aq_v}{2}\dot{Z}_i^2-\frac{1}{2a}\alpha_1(\partial Z_i)^2
-\frac{a}{2}\alpha_2Z_i^2
+\frac{1}{2a}\alpha_3(\partial V_i)(\partial Z_i)+\frac{q_t}{4a}(\partial V_i)^2\Big],
\end{aligned}
\end{equation}
where $q_v$, $\alpha_1$, $\alpha_2$ and $\alpha_3$ are given by Eqs. (4.4)-(4.7) of \cite{lavinia4}. In our case, we get
\begin{equation}\label{eqp6}
\begin{aligned}
&q_v=1,\\
&\alpha_1=1,\\
&\alpha_2=\beta_AM^2+4\beta_G\dot{H}+6\beta_GH^2,\\
&\alpha_3=-2\beta_GA_0+\dot{\phi}f_{,\phi},
\end{aligned}
\end{equation}
and $q_t$ is defined by Eq. (\ref{eq31}).\\
\noindent Varying the second-order action (\ref{eqp5}) with respect to $V_i$, we obtain the constraint equation:
\begin{equation}\label{eqp7}
\partial^2(\alpha_3Z_i+q_tV_i)=0,
\end{equation}
from this equation it is evident that $V_i=-\alpha_3Z_i/q_t$ (taking the integration constant as 0).
Replacing  this relation into Eq. (\ref{eqp5}), we obtain
\begin{equation}\label{eqp8}
S^{(2)}_v=\int{dtd^3x\sum_{i=1}^{2}\frac{aq_v}{2}\left[\dot{Z}_i^2-\frac{c_v^2}{a^2}(\partial Z_i)^2
-\frac{\alpha_2}{q_v}Z_i^2\right]},
\end{equation}
where
\begin{equation}\label{eqp9}
c_v^2=\frac{2\alpha_1q_t+\alpha_3^2}{2q_tq_v},
\end{equation}
and replacing the Eqs. (\ref{eq31}) and (\ref{eqp6}),  we obtain
\begin{equation}\label{eq36}
\begin{aligned}
c_v^2=&1+\frac{2M^2(\beta_G\beta_m+M\beta_Af_{,\phi})^2\dot{\phi}^2}{4M_{\rm Pl}^2(\beta_A M^2+6\beta_G H^2)^2-(\beta_m M-6f_{,\phi}H^2)
(M\beta_G\beta_m+2f_{,\phi}(\beta_A M^2+3\beta_G H^2))\dot{\phi}^2}.
\end{aligned}
\end{equation}
Newly, it must be fulfilled that $q_v>0$ and $c_v^2>0$ to guarantee the lack of ghost or Laplacian instabilities of vector perturbations. From Eq. (\ref{eqp6}), is evident that the first condition is satisfied.
\subsection{Scalar perturbations} 
\begin{figure*}
\centerline{\includegraphics[width=0.85\textwidth]{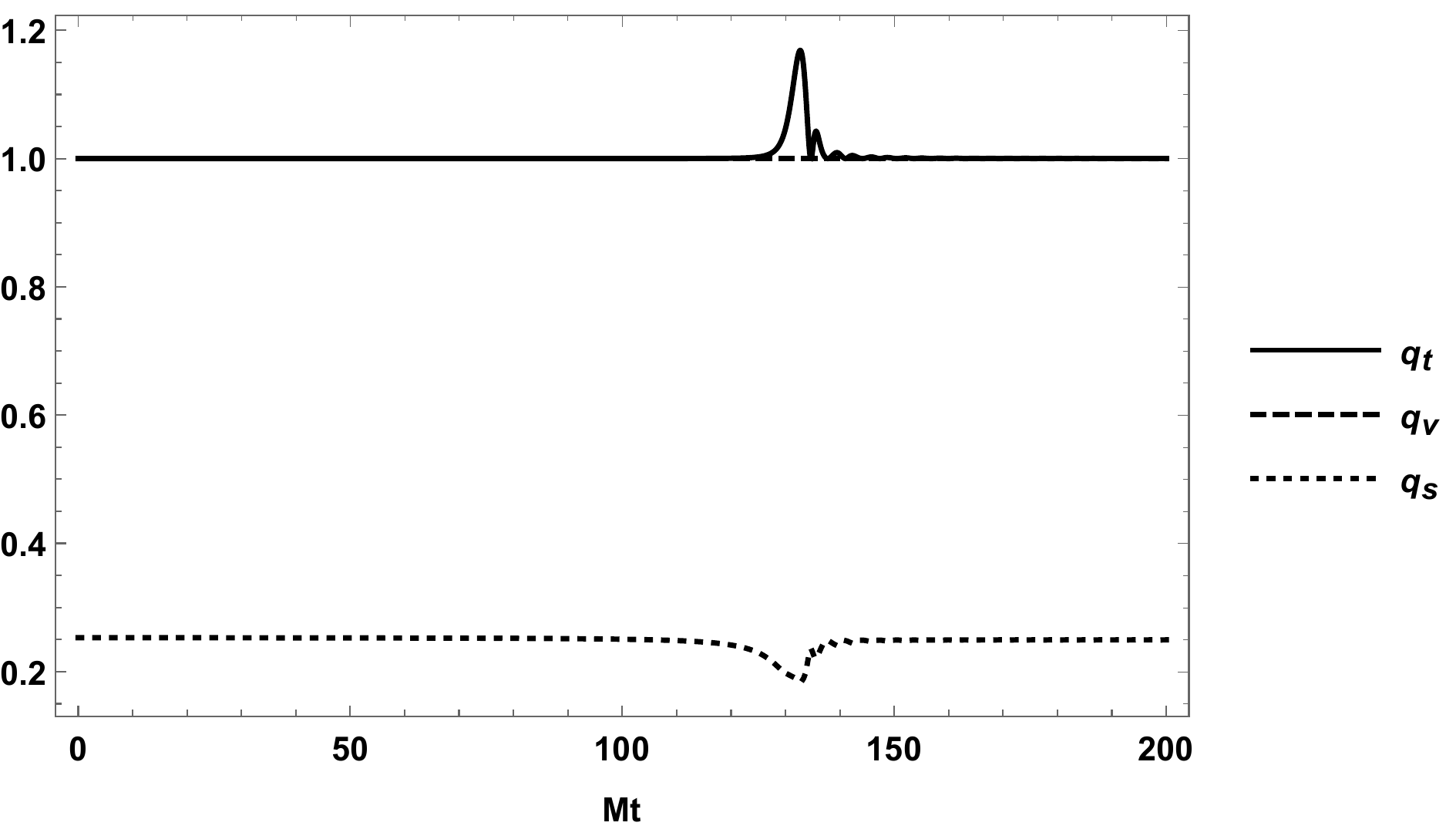}}
\caption{Evolution of $q_t$, $q_v$ and $q_s$ during  inflation. In this numerical simulation we have used the potential, the second coupling function, the values of model parameters and initial conditions as those used in Fig. \ref{fig3}
\label{fig7}}
\end{figure*}

\begin{figure*}
\centerline{\includegraphics[width=0.85\textwidth]{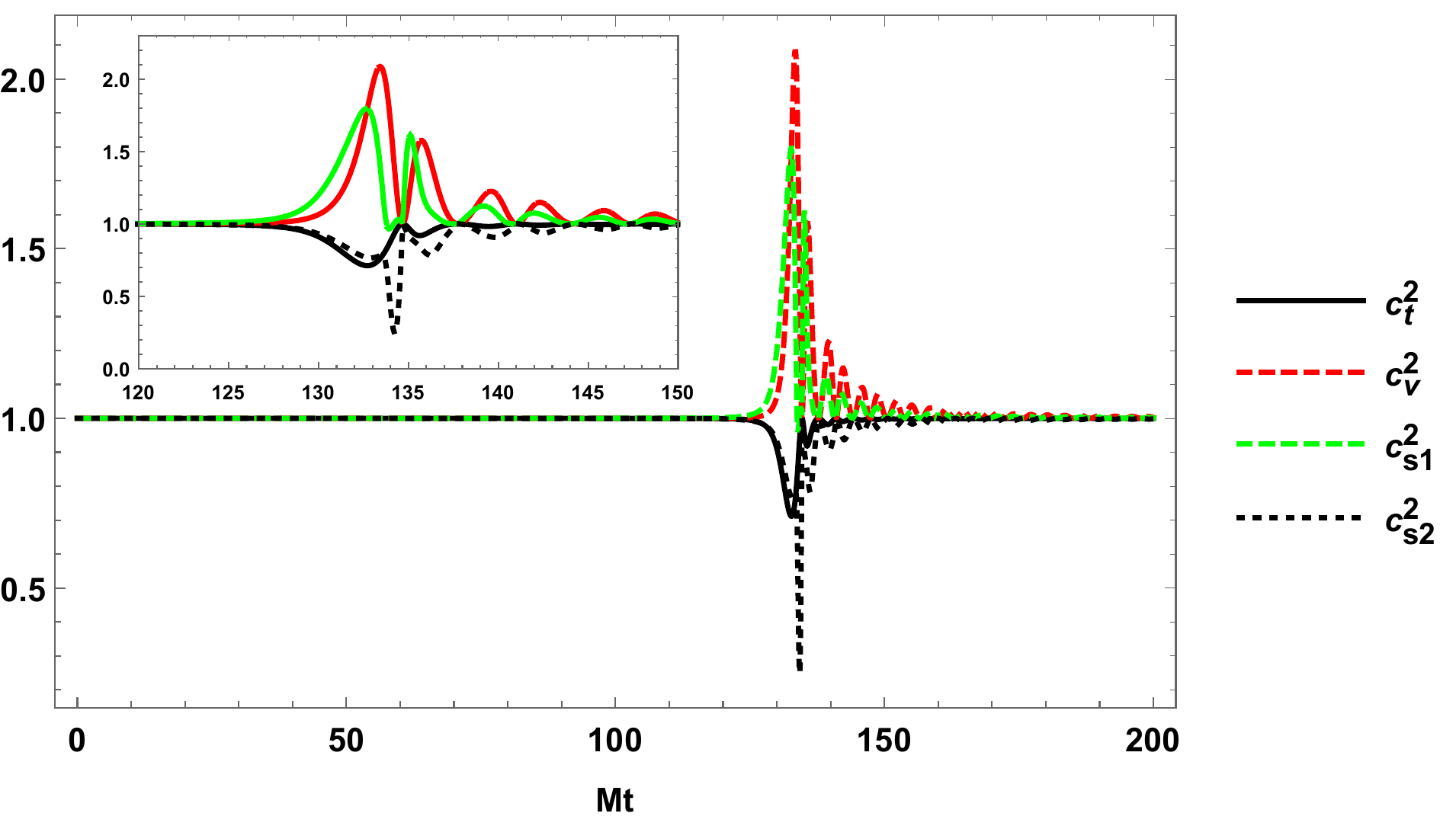}}
\caption{Evolution of $c_t^2$, $c_v^2$, $c_{s1}^2$ and $c_{s2}^2$ during  inflation. In this numerical simulation we have used the potential, the second coupling function, the values of model parameters and initial conditions as those used in Fig. \ref{fig3}\label{fig8}}
\end{figure*}

\begin{figure*}
\centerline{\includegraphics[width=0.85\textwidth]{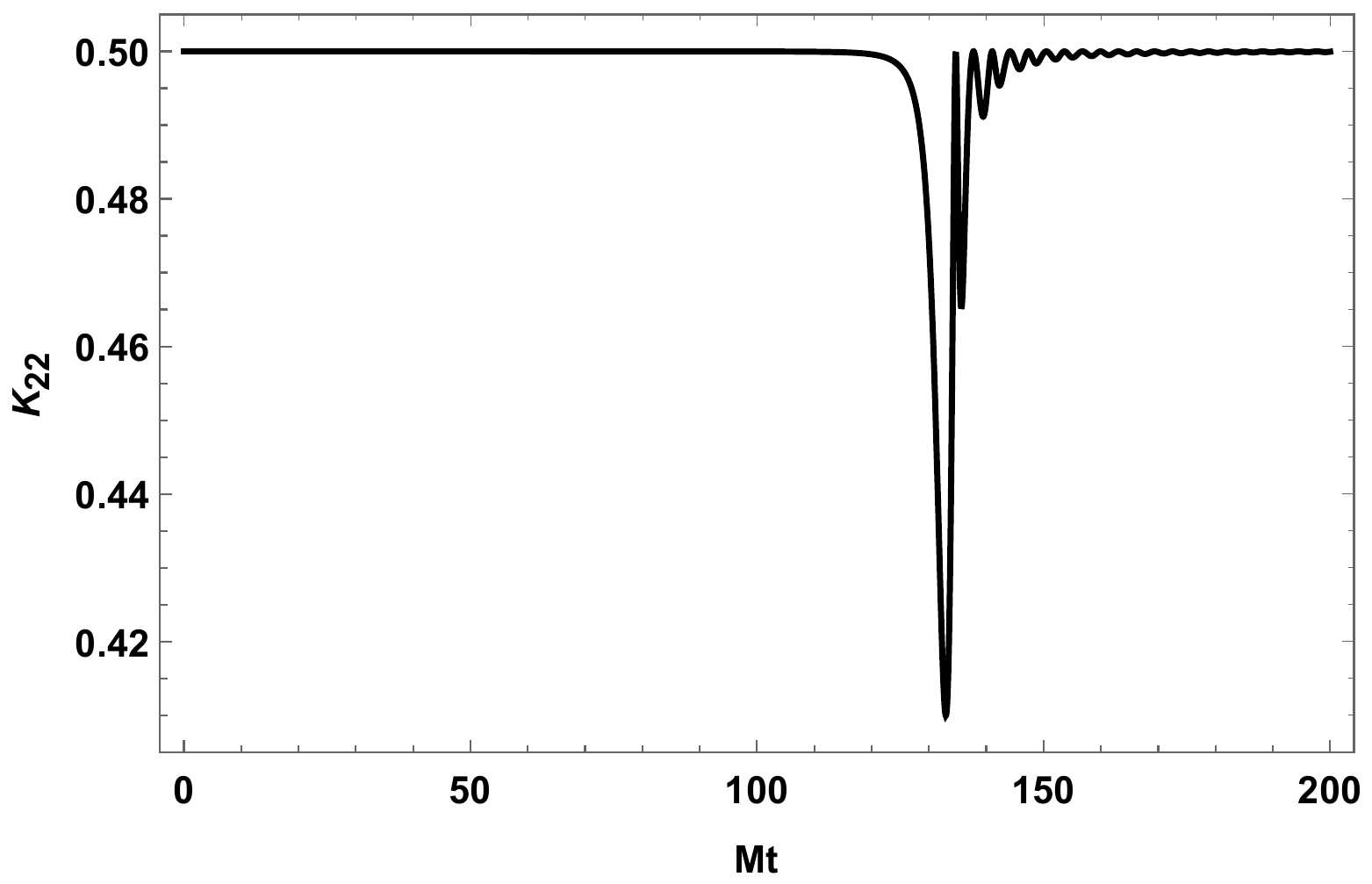}}
\caption{Evolution of the component $K_{22}$ during  inflation. In this numerical simulation we have used the potential, the second coupling function, the values of model parameters and initial conditions as those used in Fig. \ref{fig3}\label{fig9}}
\end{figure*}

\begin{figure*}
\centerline{\includegraphics[width=0.85\textwidth]{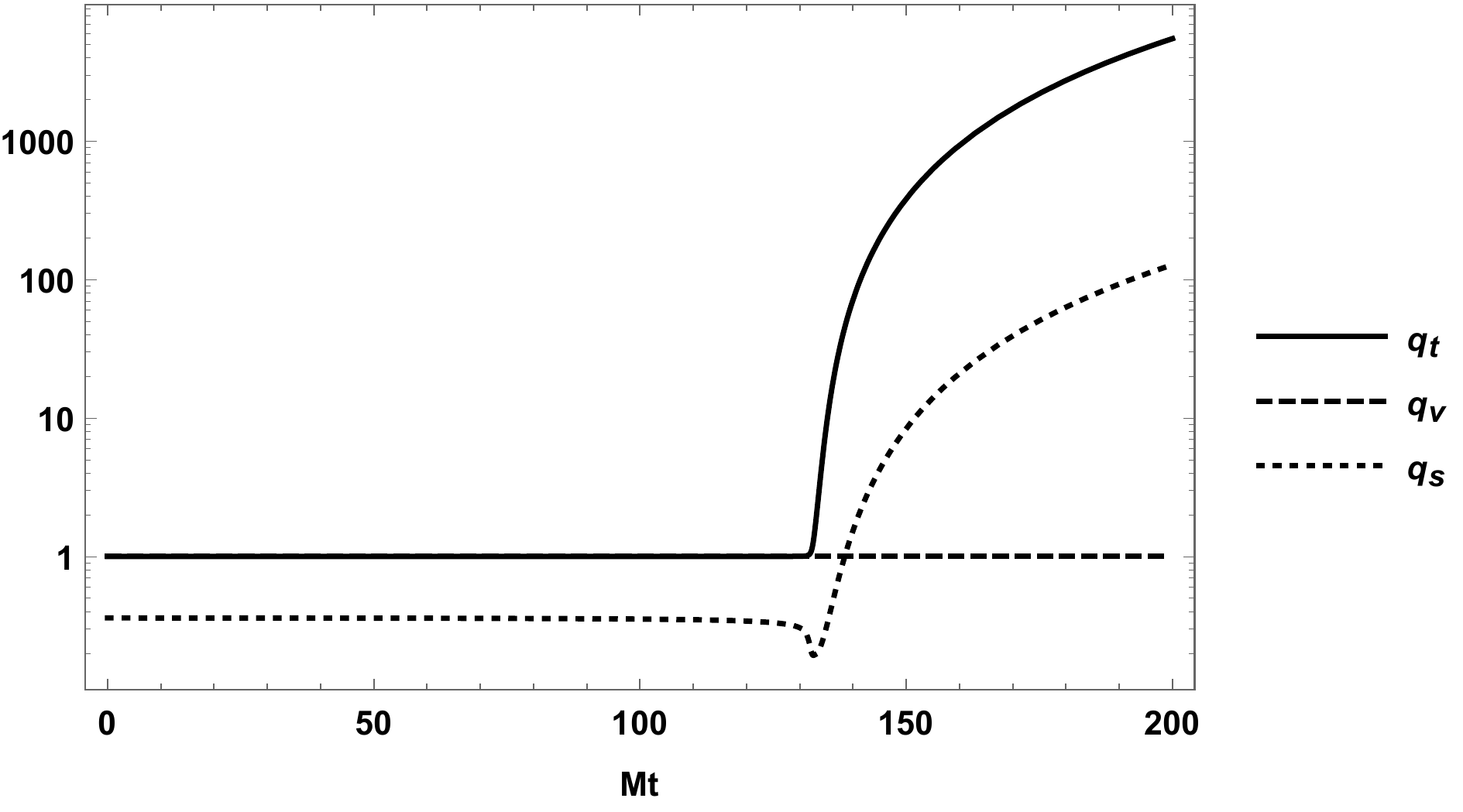}}
\caption{Evolution of $q_t$, $q_v$ and $q_s$ during  inflation. In this numerical simulation we have used the potential, the first coupling function, the values of model parameters and initial conditions as those used in Fig. \ref{fig1}
\label{fig10}}
\end{figure*}

\begin{figure*}
\centerline{\includegraphics[width=0.85\textwidth]{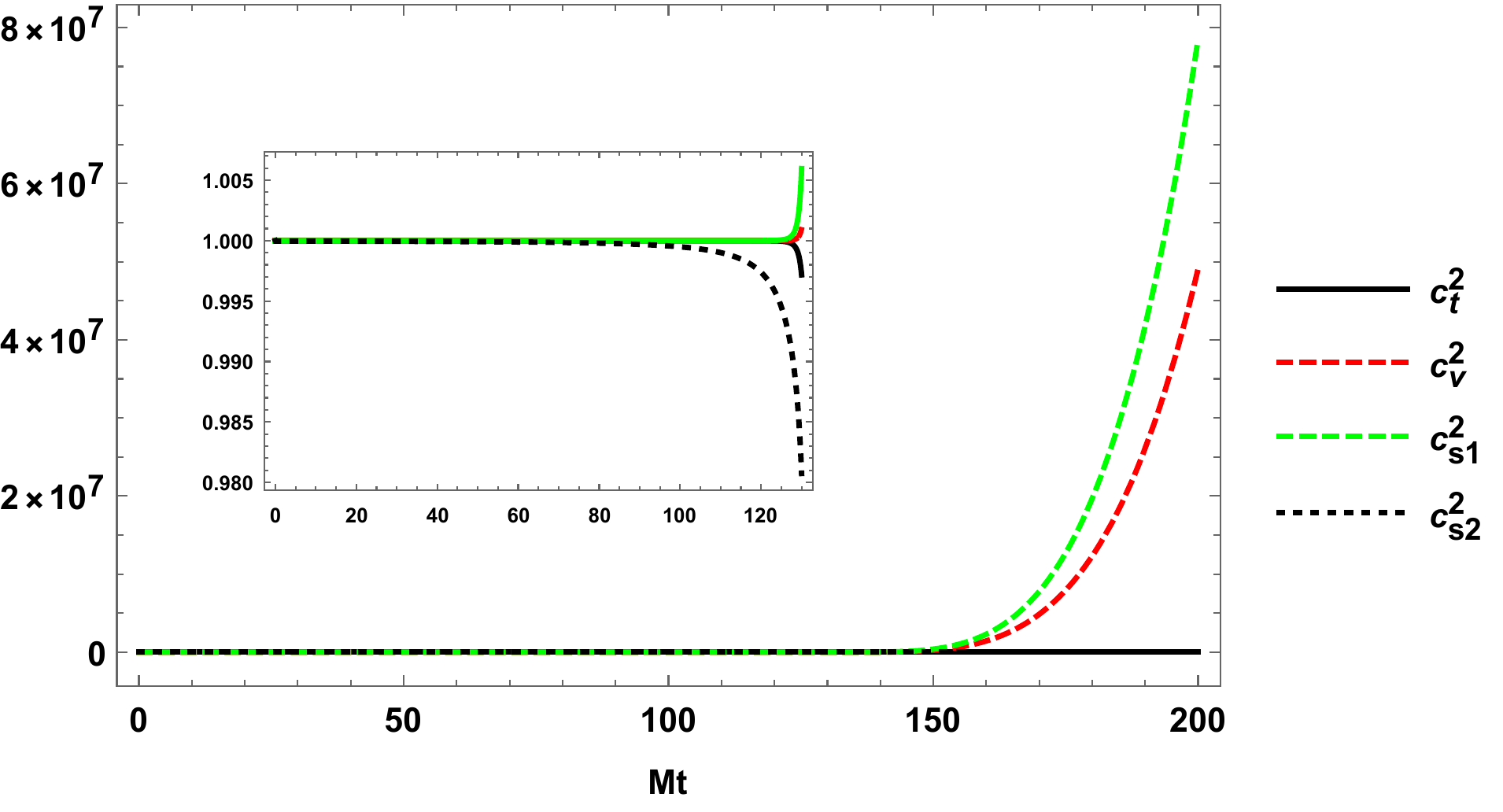}}
\caption{Evolution of $c_t^2$, $c_v^2$, $c_{s1}^2$ and $c_{s2}^2$ during  inflation. In this numerical simulation we have used the potential, the first coupling function, the values of model parameters and initial conditions as those used in Fig. \ref{fig1}\label{fig11}}
\end{figure*}

\begin{figure*}
\centerline{\includegraphics[width=0.85\textwidth]{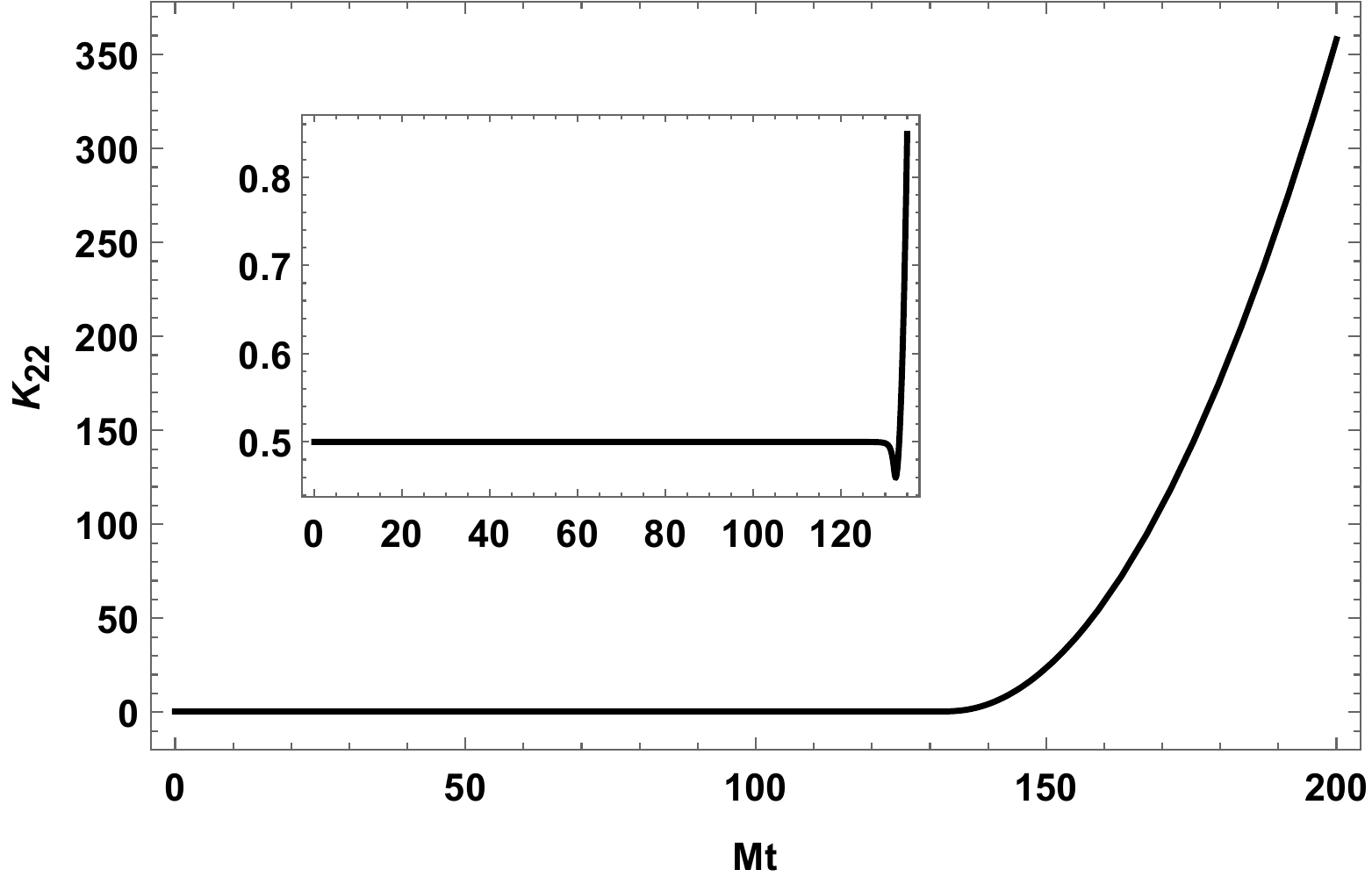}}
\caption{Evolution of the component $K_{22}$ during  inflation. In this numerical simulation we have used the potential, the first coupling function, the values of model parameters and initial conditions as those used in Fig. \ref{fig1} \label{fig12}}
\end{figure*}

For scalar perturbations, the perturbed
line element in the flat gauge:
\begin{equation}\label{eqp10}
ds^2_s=-(1+2\alpha)dt^2+2\partial_i\chi dtdx^i+a^2(t)\delta_{ij}dx^idx^j,
\end{equation}
is considered, where $\alpha$ and $\chi$ are scalar metric perturbations. The components of the vector field it can write in the form
\begin{equation}\label{eqp11}
A^0=-A_0(t)+\delta A,\ \ \ \ \ \ A_i=\partial_i\psi,
\end{equation}
where $\delta A$ is the perturbation of the temporal vector component $A^0$, and $\psi$ is the longitudinal scalar perturbation. The scalar field $\phi$ is decomposed into the background
and perturbed parts as
\begin{equation}\label{eqp12}
\phi=\phi_0(t)+\delta \phi.
\end{equation}
Expanding the action (\ref{eq1}) up to quadratic order in scalar perturbations $\alpha$, $\chi$, $\delta A$, $\psi$ and $\delta \phi$. In doing so, we use the background Eqs. (2.13) and (2.16) of \cite{lavinia4} to eliminate the terms $f_2$ and $f_{3,\phi}$. Then, the second-order action for scalar perturbations can be expressed as
\begin{equation}\label{eqp13}
S_s^{(2)}=\int{dtd^3x(\mathcal{L}^\phi_s+\mathcal{L}^{GP}_s)},
\end{equation}
where 
\begin{equation}\label{eqp14}
\begin{aligned}
\mathcal{L}^\phi_s=&a^3\Big[D_1\dot{\delta \phi}^2+D_2\frac{(\partial\delta\phi)^2}{a^2}+D_3\delta\phi^2
+\Big(D_4\dot{\delta \phi}+D_5\delta\phi+D_6\frac{\partial^2\delta\phi}{a^2}\Big)\alpha-(D_6\dot{\delta\phi}
-D_7\delta\phi)\frac{\partial^2\chi}{a^2}\\
&+(D_8\dot{\delta\phi}
+D_9\delta\phi)\delta A+D_{10}\delta\phi\frac{\partial^2\psi}{a^2}\Big],
\end{aligned}
\end{equation}
and
\begin{equation}\label{eqp15}
\begin{aligned}
\mathcal{L}^{GP}_s=&a^3\Big[\Big(w_1\alpha-w_2\frac{\delta A}{A_0}\Big)\frac{\partial^2\chi}{a^2}-w_3\frac{(\partial\alpha)^2}{a^2}+w_4\alpha^2
-\Big(w_3\frac{\partial^2\delta A}{a^2A_0}-w_8\frac{\delta A}{A_0}+w_3\frac{\partial^2\dot{\psi}}{a^2A_0}
+w_6\frac{\partial^2\psi}{a^2}\Big)\alpha\\
&-w_3\frac{(\partial\delta A)^2}{4a^2A_0^2}+w_5\frac{\delta A^2}{A_0^2}
+\{w_3\dot{\psi}-(w_2-A_0w_6)\psi\}\frac{\partial^2\delta A}{2a^2A_0^2}
-w_3\frac{(\partial\dot{\psi})^2}{4a^2A_0^2}+w_7\frac{(\partial\psi)^2}{2a^2}\Big],
\end{aligned}
\end{equation}
where $GP$ means Generalized Proca thaories and the coefficients $D_{1,\ldots,10}$ and $w_{1,\ldots,8}$ were defined in the appendix of \cite{lavinia4}.  In our case, these coefficients are given by
\begin{equation}\label{eq37}
\begin{aligned}
&D_1=\frac{1}{2},\\
&D_2=-\frac{1}{2},\\
&D_3=\frac{9}{2}H^3A_0f_{,\phi\phi}+\frac{1}{2}V_{,\phi\phi},\\
&D_4=-\dot{\phi}+3H^2A_0f_{,\phi}+\frac{1}{2}\beta_m MA_0,\\
&D_5=3H^2\dot{\phi}A_0f_{,\phi\phi}-V_{,\phi},\\
&D_6=-2HA_0f_{,\phi},\\
&D_7=-3H^2\dot{\phi}A_0f_{,\phi}+2H\dot{\phi}A_0f_{,\phi\phi}+\dot{\phi}+\frac{1}{2}\beta_m MA_0,\\
&D_8=-\frac{2\dot{\phi}D_1+D_4+3HD_6}{A_0},\\
&D_9=-\frac{D_5}{A_0}+6H^2\dot{\phi}f_{,\phi\phi}-\frac{V_{,\phi}}{A_0},\\
&D_{10}=-2\dot{H}f_{,\phi}-3H^2f_{,\phi}+\frac{1}{2}\beta_m M
\end{aligned}
\end{equation}
and
\begin{equation}\label{eq38}
\begin{aligned}
&w_1=-2H\dot{\phi}A_0f_{,\phi}-2HM_{\rm Pl}^2-2\beta_GHA_0^2,\\
&w_2=w_1+2Hq_t-\dot{\phi}D_6,\\
&w_3=-2A_0^2q_v,\\
&w_4=-3H^2M_{\rm Pl}^2+3H^2\beta_GA_0^2+\frac{3}{2}H^2\dot{\phi}A_0f_{,\phi}+\frac{1}{2}\dot{\phi}^2
-\frac{3}{4}\beta_m M\dot{\phi}A_0,\\
&w_5=w_4-\frac{3H(w_1+w_2)}{2}-\frac{1}{2}\dot{\phi}^2+\frac{1}{2}\beta_m M\dot{\phi}A_0,\\
&w_6=-\frac{2w_1-w_2-2\dot{\phi}D_6+4HM_{\rm Pl}^2+4\beta_GHA_0^2}{A_0},\\
&w_7=-4\beta_G \dot{H}-\frac{3H^2\dot{\phi}f_{,\phi}}{A_0}+\frac{\beta_m M\dot{\phi}}{2A_0},\\
&w_8=3Hw_1-2w_4+\dot{\phi}D_4.
\end{aligned}
\end{equation}
From Eqs. (\ref{eqp14}) and (\ref{eqp15}) it is clear that  the fields $\alpha$, $\chi$ and $\delta A$ are nondynamical. Therefore, varying the action (\ref{eqp13}) with respect to $\alpha$, $\chi$ and $\delta A$, we obtain the three constraint equations in Fourier space:
\begin{equation}\label{eqp16}
\begin{aligned}
&D_4\dot{\delta\phi}+D_5\delta\phi+2w_4\alpha+w_8\frac{\delta A}{A_0}
+\frac{k^2}{a^2}\Big(w_3\frac{\dot{\psi}}{A_0}+w_6\psi-D_6\delta\phi-2w_3\alpha-w_1\chi+w_3\frac{\delta A}{A_0}\Big)=0,
\end{aligned}
\end{equation}
\begin{equation}\label{eqp17}
D_6\dot{\delta\phi}-D_7\delta\phi-w_1\alpha+w_2\frac{\delta A}{A_0}=0,
\end{equation}
\begin{equation}\label{eqp18}
\begin{aligned}
&D_8\dot{\delta\phi}+D_9\delta\phi+w_8\frac{\alpha}{A_0}+2w_5\frac{\delta A}{A^2_0}-\frac{k^2}{a^2}\frac{1}{A_0}
\Big(\frac{w_3}{2}\frac{\dot{\psi}}{A_0}+\frac{A_0w_6-w_2}{2}\frac{\psi}{A_0}-w_3\alpha-w_2\chi+\frac{w_3}{2}\frac{\delta A}{A_0}\Big)=0,
\end{aligned}
\end{equation}
we solve Eqs. (\ref{eqp16})-(\ref{eqp18}) for $\alpha$, $\chi$ and $\delta A$, and eliminate these variables from the action (\ref{eqp13}). Then, the second-order action of scalar perturbations can be expressed in the form
\begin{equation}\label{eqp19}
S_v^{(2)}=\int{dtd^3xa^3\left(\dot{\vec{\chi}}^t\mathbf{K}\dot{\vec{\chi}}-\frac{k^2}{a^2}\vec{\chi}^t\mathbf{G}\vec{\chi}-
\vec{\chi}^t\mathbf{M}\vec{\chi}-\vec{\chi}^t\mathbf{B}\dot{\vec{\chi}}\right)},
\end{equation}
where $\mathbf{K}$, $\mathbf{G}$, $\mathbf{M}$ and $\mathbf{B}$ are $2\times 2$ matrices, and $\vec{\chi}$ is defined by
\begin{equation}\label{eqp20}
\vec{\chi}^t=(\psi,\delta\phi).
\end{equation}
In the small-scale limit, the leading-order contributions to the matrix $\mathbf{M}$ do not contain the $k^2$ terms. We shift the $k^2$ terms
appearing in $\mathbf{B}$ to the matrix components of $\mathbf{G}$ after integrating them by parts. For $k\rightarrow \infty$, the components of $\mathbf{K}$ and $\mathbf{G}$ are given, respectively, by
\begin{equation}\label{eqp21}
\begin{aligned}
&K_{11}=\frac{w_1^2w_5+w_2^2w_4+w_1w_2w_8}{A_0^2(w_1-2w_2)^2},\\
&K_{22}=D_1+\frac{D_6}{w_1-2w_2}\Big(D_4+\frac{w_4+4w_5+2w_8}{w_1-2w_2}D_6+2A_0D_8\Big),\\
&K_{12}=K_{21}=-\frac{1}{2A_0(w_1-2w_2)}
\Big[w_2D_4+\frac{w_1(4w_5+w_8)+2w_2(w_4+w_8)}{w_1-2w_2}D_6+A_0w_1D_8\Big],
\end{aligned}
\end{equation}
and
\begin{equation}\label{eqp22}
\begin{aligned}
&G_{11}=\dot{E}_1+HE_1-\frac{4A_0^2}{w_3}E_1^2-\frac{w_7}{2},\\
&G_{22}=\dot{E}_2+HE_2-\frac{2A_0}{w_2}D_7E_3-\frac{4A_0^2}{w_3}E_3^2-D_2,\\
&G_{12}=G_{21}=\dot{E}_3+HE_3-\frac{4A_0^2}{w_3}E_1E_3
+\frac{w_2}{2A_0(w_1-2w_2)}D_7+\frac{D_{10}}{2},
\end{aligned}
\end{equation}
where $E_1$, $E_2$ and $E_3$ are defined as
\begin{equation}\label{eqp23}
\begin{aligned}
&E_1=\frac{w_6}{4A_0}-\frac{w_1w2}{4A_0^2(w_1-2w_2)},\\
&E_2=-\frac{D_6^2}{2(w_1-2w_2)},\\
&E_3=\frac{w_2D_6}{2A_0(w_1-2w_2)}.
\end{aligned}
\end{equation}
In our case, the explicit form for the components  Eqs. (\ref{eqp21}) and (\ref{eqp22})  are obtained replacing $D_{1,\ldots,10}$ and $w_{1,\ldots,8}$ given by Eqs. (\ref{eq37}) and (\ref{eq38}), but the resulting expressions are too long to be written here.\\
\noindent In order to ensure the absence of scalar ghosts, the kinetic matrix $\mathbf{K}$ must be positive definite. In other words, the determinants of principal submatrices of $\mathbf{K}$ need to be positive. Thus, we require the following two no-ghost conditions:
\begin{equation}\label{eqp24}
K_{11}>0\ \ \ \ \text{or}\ \ \ \ K_{22}>0,
\end{equation}
\begin{equation}\label{eqp25}
q_s\equiv K_{11}K_{22}-K_{12}^2.
\end{equation}
In the small-scale limit, the dispersion relation following from the action (\ref{eqp19}) with frequency $\omega$ is given by
\begin{equation}\label{eqp26}
\text{det}\left(\omega^2\mathbf{K}-\frac{k^2}{a^2}\mathbf{G}\right)=0,
\end{equation}
Introducing the scalar sound speed $c_s$ as $\omega^2=c_s^2k^2/a^2$, the above dispersion relation leads to the two scalar propagation speed squares:
\begin{equation}\label{eqp27}
\begin{aligned}
c_{s1}^2&=\frac{K_{11}G_{22}+K_{22}G_{11}-2K_{12}G_{12}+\sqrt{(K_{11}G_{22}+K_{22}G_{11}
-2K_{12}G_{12})^2-4(K_{11}K_{22}-K_{12}^2)(G_{11}G_{22}-G_{12}^2)}}
{2(K_{11}K_{22}-K_{12}^2)},
\end{aligned}
\end{equation}
\vspace{0.5cm}
\begin{equation}\label{eqp28}
\begin{aligned}
c_{s2}^2&=\frac{K_{11}G_{22}+K_{22}G_{11}-2K_{12}G_{12}-\sqrt{(K_{11}G_{22}+K_{22}G_{11}
-2K_{12}G_{12})^2-4(K_{11}K_{22}-K_{12}^2)(G_{11}G_{22}-G_{12}^2)}}
{2(K_{11}K_{22}-K_{12}^2)}.
\end{aligned}
\end{equation}
For the absence of Laplacian instabilities, we require that
\begin{equation}\label{eqp29}
c_{s1}^2>0,\ \ \ \ \ c_{s2}^2>0.
\end{equation}
In order for the system to be completely stable against ghosts and Laplacian instabilities, all
the dynamical degrees of freedom must simultaneously satisfy both of the conditions of no-ghosts
and positive squared sound speeds, however the expressions obtained here for $q_t$, $c_t^2$, $q_v$, $c_v^2$, $K_{22}$, $q_s$, $c_{s1}^2$ $c_{s2}^2$  are in terms of complicated combinations, which include several model parameters and other quantities as $H$, $f_{,\phi}$  and the field $\phi$ and its derivative, then is not easy to write explicit no-ghosts and  stability conditions. Therefore, in this work we will verify numerically these conditions taking into account particular choices for the model parameters, the potential $V(\phi)$, the coupling functions $f(\phi)$ and the  initial conditions
used in the previous section. Thereby, in Figs. \ref{fig7}, \ref{fig8}, \ref{fig9},  \ref{fig10}, \ref{fig11} and \ref{fig12} we can see that for our model the no-ghosts and stability conditions during  inflation are satisfied, i.e. there are no ghosts and Laplacian instabilities of tensor, vector, and scalar
perturbations. In Figs. \ref{fig8} and \ref{fig11} we observe that the given quantities are nearly 1 during inflation (in this work, we have used the natural units in which the speed of light $c$ is equivalent to 1) and in Fig. \ref{fig9} and \ref{fig12}
we see that $K_{22}$ is about  1/2 during inflation. However,  at the end of inflation  these figures  show a large deviation  which is induced by the time variation of $\phi$ (besides, in Fig. \ref{fig1} we can see that $\dot{\phi}$ reaches a minimum around the end of inflation). This same behavior is observed in Fig. \ref{fig3} too. Moreover, in this context, the general condition is that these quantities must be positive to guarantee the absence of instabilities within the model, since in this regimen (inflation), until now, there are no constraints imposed by the observations in early times, like those imposed by the gravitational waves in late times, $z<0.09$,   (the speed of gravitational waves $c_t$ is the same as the speed of light, which places tight constraints on dark energy models constructed in the framework of modified gravitational theories). In this sense, this restriction does not apply to high redshift values (inflation), thus, the model (scalar–vector-tensor theories) under consideration in this work could be used in an inflationary context. Finally, from the above, we can say that using the second coupling function we obtain better results than those obtained using the first coupling function (since in Fig. \ref{fig11} the exponential growth of  $c_v^2$ and $c_{s1}^2$ after inflation could be problematic).

\section{Conclusions}\label{conclus} 
In this work,  we have studied  inflation in a particular scalar-vector-tensor theory of gravitation without the $U(1)$ gauge symmetry. This model was constructed from the more general action given by Eqs. (\ref{eq1}) and (\ref{eq2}) using certain specific choices  for the Lagrangians and the coupling functions (see Eq. (\ref{eq4})). From this model, we obtained the general background equations of motion on the flat FLRW spacetime and using concrete choices for the potential (see Eq. (\ref{eq30})), the coupling functions (see Eq. (\ref{eq30'})), suitable dimensionless coupling constants  and initial conditions, was possible to verify numerically that this model of inflation is viable. In this way,  in Fig. \ref{fig1} we plotted the evolution of $\dot{\phi}$  and $A_0$  during inflation and reheating, and from it, is clear that  $\dot{\phi}$  and $A_0$ display a slow evolution during inflation. Further, the fields decay faster during reheating. 
\noindent  In Fig. \ref{fig2} the evolution of the number of $e$-foldings $N$ was plotted and we see that in this case, it is possible to obtain a suitable amount of $e$-foldings ($N\gtrsim 60$) for sufficient inflation. Moreover, $N\approx 63$ (for $\omega=1\,M^4$) at the end of inflation. Using the second coupling function, in Fig. \ref{fig3}, we can see  a similar behavior during inflation to the shown in Fig. \ref{fig1}, but during reheating the
fields oscillate and the amplitude of $A_0$ decreases faster than the amplitude of  $\dot{\phi}$. Furthermore, in Fig. \ref{fig4} is clear
that for the second choice of the coupling function (see Eq. (\ref{eq30'})), we got a similar amount of $e$-foldings  as that obtained 
using the first coupling function. In this case, $N\approx 64$ (for $\xi=4.5$) at the end of inflation. Therefore,  we can say  that the introduction of the coupling function $f(\phi)$ in our model of inflation, allows us to reach a suitable amount of $e$-foldings $N$ for sufficient inflation. This is a remarkable result, since without the coupling function contribution,  the amount of $e$-foldings is
smaller at the end of inflation, as has been demonstrated in \cite{lavinia4}. In Fig. \ref{fig5} we plotted the evolution of
the new slow-roll parameters $\delta_1$ and $\delta_2$ which are defined by Eq. (\ref{eq25}) and from it, is evident that the slow-roll parameter parameters satisfy the conditions $|\delta_1|\ll 1$ and $|\delta_2|\ll 1$ during  inflation. Also, we have shown that the numerical inflationary solution obtained here  is an attractor solution (see Fig \ref{fig6}). 
Finally, the no-ghosts and stability conditions that the model during inflation must satisfy, i.e., absence of ghosts and Laplacian instabilities of linear cosmological perturbations were obtained. For do this, the general expressions for tensor, vector and scalar perturbations obtained in \cite{lavinia4} were used,  reproducing literally the main results associated with the general stability analysis which was performed in the sections \textbf{III}, \textbf{IV} and \textbf{V} of \cite{lavinia4}, and then we applied this general formalism to our model. In this sense, the main expressions which must be positive to guarantee absence of ghosts and Laplacian instabilities were obtained, and them are: $q_t$, $c_t^2$, $q_v$, $c_v^2$, $K_{22}$, $q_s$, $c_{s1}^2$ $c_{s2}^2$ (see Eqs. (\ref{eq31}), (\ref{eq32}), (\ref{eqp6}), (\ref{eq36}), (\ref{eqp24}), (\ref{eqp25}), (\ref{eqp27}) and (\ref{eqp28})).
Additionally, the positivity condition which these quantities must be satisfied were verified numerically too (see Figs. \ref{fig7}-\ref{fig12}).\\
\noindent We must emphasize that the main goal of this work  was devoted to the demonstration of a viable model of inflation
that was healthy at the theoretical level,  which was successfully achieved. Phenomenological consequences of this pathology-free model is therefore of interest for further studies. The observational bounds on the scalar spectral index and the tensor-to-scalar ratio can constrain/falsify this model, but that kind of analysis is beyond the scope of this work and could be addressed later.\\

\noindent \textbf{Acknowledgements}
This work was supported by  Patrimonio Aut\'onomo-Fondo Nacional de Financiamiento para la Ciencia, la Tecnolog\'ia y la Innovaci\'on Francisco Jos\'e de Caldas (MINCIENCIAS-COLOMBIA) Grant No. 110685269447 RC-80740-465-2020, projects 69723 and 69553.

\end{document}